\documentclass[aps,reprint,twocolumn,superscriptaddress,10pt]{revtex4-2}
\usepackage[dvipdfmx]{graphicx}

\usepackage{siunitx}
\usepackage{amsmath}
\usepackage{amssymb}
\usepackage{bm}
\usepackage{graphicx}
\usepackage{dcolumn}
\usepackage{euscript}
\usepackage{multirow}
\usepackage{color}
\usepackage{textcomp}
\usepackage{gensymb}
\usepackage{float}
\usepackage{enumitem}
\usepackage{xcolor}
\graphicspath{{./figs/}}
\usepackage[outdir=./figs/]{epstopdf}

\newcommand{\rfig}[1]{Fig.~\ref{#1}}
\newcommand{\rFig}[1]{Figure~\ref{#1}}
\newcommand{\rtbl}[1]{Table~\ref{#1}}

\newcommand{\sfig}[1]{Fig.S{#1}}
\newcommand{\stbl}[1]{Table.S{#1}}

\newcommand\etal{\textit{et al.}}
\newcommand{\mub}{\mu_\text{B}}
\newcommand{\mspin}{m_{s}}
\newcommand{\morb}{m_{l}}
\newcommand\tc{T_\text{C}}

\newcommand{\cgt}{\text{Cr}\text{Ge}\text{Te}_3}

\newcommand{\cri}{\text{CrI}_3}
\newcommand{\fgt}{\text{Fe}_3\text{GeTe}_2}

\begin{document}
\title{X-ray spectra in magnetic van der Waals materials $\fgt$,
  $\cri$, and $\cgt$: a first-principles study}

\author{Y. Lee}
\affiliation{Ames Laboratory, U.S.~Department of Energy, Ames, Iowa 50011}
\author{V. N. Antonov}
\affiliation{G. V. Kurdyumov Institute for Metal Physics of the N.A.S. of Ukraine, 36 Academician Vernadsky Boulevard, UA-03142 Kyiv, Ukraine}
\author{B. N. Harmon}
\affiliation{Ames Laboratory, U.S.~Department of Energy, Ames, Iowa 50011}
\author{Liqin Ke}
\affiliation{Ames Laboratory, U.S.~Department of Energy, Ames, Iowa 50011}


\date{\today} 

\begin{abstract}
  Using density functional theory (DFT) methods, we have calculated
  X-ray absorption spectroscopy (XAS) and X-ray circular dichroism
  (XMCD) spectra in bulk and thin films of $\fgt$, $\cri$, and $\cgt$.
  DFT+$U$ methods are employed for better handling of correlation
  effects of 3$d$ electrons of transition metals.  We discuss
  relations between the density of states, radial matrix elements, and
  the corresponding spectra.  By comparing the calculated spectra with
  previously measured spectra, we discuss the reliability of DFT+$U$
  methods to describe the electronic structures of these materials and
  determine the corresponding optimal $U$ and $J$ parameters.
\end{abstract}

\maketitle

\section{Introduction}

Graphene is an attractive material because of its novel electronic structure and its
potential for practical
applications~\cite{allen2010chemicalreview,choi2010critical,acik2011,coros2019}.
However, the lack of intrinsic magnetism is a limitation for some
device applications.  Nevertheless, considerable effort has been
expended to induce and control magnetism in
graphene~\cite{yazyev201271,kan2008ws} and separately to discover new
two-dimensional (2D) materials that have intrinsic magnetism.  Recent
unprecedented experimental realization of magnetic 2D van
der Waals (vdW) materials~\cite{gong2017n,huang2017n} has created
great excitement and determination to explore magnetism in these new
2D materials.

Among the magnetic 2D vdW (m2DvdW) materials, $\cri$, $\cgt$ and $\fgt$ are the most
intensively investigated ferromagnetic materials to understand
fundamental physics such as the mechanism of magnetic ordering, the
exchange interaction and the magnetocrystalline
anisotropy as well as to improve desired properties by controlling external conditions such
as electric and magnetic fields, strain, pressure, and doping~\cite{jang2019prm,lado20172m,li2019pressure,subhan2019pressure,mondal2019pressure,jiang2018doping,wang2016doping,xu2020doping,su2019efield}.
Furthermore, since it is easy to separate vdW materials into thin
layers and to over-lay different materials, the creation and
exploration of composite thin layers has expanded the study of a
wide variety of new interfaces and composite layered
materials~\cite{zutic2019sciencedirect,liu2019nature,kang2013apl,shang2020acsanm,dayen2020apr,gibertini2019naturenano,shi2018sst}.

To investigate magnetic properties of heterostructures and alloys
consisting of different magnetic atoms, it is beneficial to separate
contributions from different atoms instead of averaging or summing
them.  X-ray Magnetic Circular Dichroism (XMCD) is useful for
understanding the roles of individual magnetic atoms responsible for
collective magnetic properties such as the magnetic ordering in
interfaces of heterostructures. For instance, Liu
$\etal$~\cite{liu2019nsr} performed XMCD measurements on a
superlattice structure of $\fgt$ and CrSb; and discovered that $\tc$
of $\fgt$ can be significantly enhanced because of the interfacial
ferromagnetic coupling. Also, Burn $\etal$
\cite{burn2019scientificreports} were able to characterize magnetic
properties of the layers of Cr$_{2}$Te$_{3}$ thin films and
Cr$_{2}$Te$_{3}$/Cr:Sb$_{2}$Te$_{3}$ (Cr doped Sb$_{2}$Te$_{3}$)
heterostructures by XMCD measurements.

X-ray absorption spectroscopy (XAS) and XMCD have become essential
tools for investigating magnetic properties of ferromagnetic alloys,
surfaces and interfaces as well as magnetic bulk
compounds because of the chemical and orbital selectivity of X-ray spectra ~\cite{stohr2006magnetism,wende2004rpp,zhang2019acsnano,aruta2009prb,srivastava1998prb}.
These tools are able to resolve the total magnetic moment into the
orbital and spin contributions by using sum rules.  And they can
assess magnetic information of each element within compounds and
interfacial structures.  Moreover, since these techniques involve
electron transitions from well defined core states to (at times
complex) conduction states, the spectra provide knowledge of the
composition of the density of states (DOS) and the orbital character
of the conduction bands.

Since the magnetic moment is a number that involves integration, it is
not easy to pinpoint the origin or details of any differences of states by comparing their magnetic moments.  On the other hand a comparison of DOSs can give information about the differences.  In the experiment, optical spectra give energy-resolved
information and can be compared with DOS($E$).  In XAS/XMCD, since
initial states are well defined atomic core states, which are not
affected by surroundings, it is useful to obtain information of final
conduction states.  Comparison of theoretical spectra which are
calculated with various methods and experimental XAS/XMCD spectra help
to decide the more suitable theoretical method for further work.  It is
possible to associate key spectra structures with calculated DOS and
obtain electronic structure information.

There are a number of reports for XMCD measurements for bulk m2DvdW
materials such as $\fgt$~\cite{zhu2016prb,park2020nanoletter},
$\cri$~\cite{frisk2018materialletters,kim2019prl} and
$\cgt$~\cite{watson2020prb}.
For $\fgt$, the XAS and XMCD spectra of Fe L$_{3,2}$ edges have been
measured~\cite{zhu2016prb,park2020nanoletter}.  Zhu
$\etal$~\cite{zhu2016prb} extracted the orbital moment of Fe atoms
from the spectra and found good agreement with their calculations and Park $\etal$~\cite{park2020nanoletter} showed that the ratio
between orbital and spin moments of Fe atoms ($\sim 0.092$) in $\fgt$
is much higher than in elemental Fe ($\sim$0.04).  Li
$\etal$\cite{li2018nanoletter} measured the temperature-dependent XMCD
spectra to study the magnetic ordering of $\fgt$ and found a $\tc$ of
\SI{230}{\K}, which agrees with the SQUID result.

For $\cri$, Frisk $\etal$~\cite{frisk2018materialletters} measured the
Cr L$_{3,2}$ edge spectra and calculated the spectra using atomic
multiplet theory~\cite{thole1985prb}.
Using XMCD spectra, Kim $\etal$~\cite{kim2019prl} estimated in-plane and
out-of-plane orbital moments ($\morb$) and concluded that the
anisotropy of magnetic orbital moments in $\cri$ is negligible, unlike
in $\cgt$.
For $\cgt$, Waston $\etal$~\cite{watson2020prb} performed XAS and XMCD
measurements and atomic multiplet calculations for the Cr L$_{3,2}$
edge spectra of $\cgt$.
Along with angle-resolved photoemission spectroscopy (ARPES)
measurements, they identified covalent bonding states and suggested
these states as the primary driver of the ferromagnetic ordering of
$\cgt$.

Most theoretical studies of the X-ray spectra in m2DvdW materials so
far were based on atomic multiplet theory or cluster
models~\cite{de2008core}, which employ adjustable parameters to
describe the electronic structure and X-ray spectra.  Though the
reported theoretical spectra line shapes may show good agreement with
experiments, it is hard to interpret the electronic structure in a
comprehensive way because they rely on selected parameters.  Atomic
multiplet theory allows an easy incorporation of many-body effects but
is not reliable for interpreting solid-state-like effects.  Therefore,
solid state first-principles calculations are useful for an integrated
understanding of the system while the atomic and empirical approaches
provide complementary information.

In this study, we calculate and discuss the electronic structure and
X-ray spectra of $\fgt$, $\cri$, and $\cgt$.  We performed
calculations not only with bulk structures but also with thin-film
structures since most of reported X-ray spectra
measurements~\cite{zhu2016prb, frisk2018materialletters,kim2019prl}
were done in total-electron-yield (TEY) mode which is surface
sensitive and the probing depth of TEY is about 3-10 \textit{nm},
depending on materials~\cite{de2008core}.  Since the
\textit{c}-lattice parameter is about or less than 2 \textit{nm} ,
X-ray spectra in TEY mode may not yield bulk structure information for
$\cri$, $\cgt$ and $\fgt$.  Furthermore, the m2DvdW are attractive
because of their thin-layer character.  Therefore it is interesting to
investigate the X-ray spectra of thin films (layers).  We show that
X-ray spectra are useful for determining the unoccupied DOS through
comparison between the partial DOSs and spectral line shape.
Electronic structures are described within density functional theory
(DFT) and DFT+$U$. By comparison between the theoretical and experimental spectra, 
we determine the optimal $U$ and $J$ values that are able to describe the electronic structures of these materials more satisfactorily within the DFT+$U$ framework.

\section{Theory and computational details}

\subsection{X-ray magnetic circular dichroism.} 

Magneto-optical (MO) effects refer to various changes in the
polarization state of light upon interaction with materials possessing
a net magnetic moment, including rotation of the plane of linearly
polarized light (Faraday, Kerr rotation), and the complementary
differential absorption of left and right circularly polarized light
(circular dichroism). In the near visible spectral range these effects
result from excitation of electrons in the conduction band. Near x-ray
absorption edges, or resonances, magneto-optical effects can be
enhanced by transitions from well-defined atomic core levels to
empty valence or conduction states.

Within the one-particle approximation, the absorption coefficient
$\mu^{\lambda}_j (\omega)$ for incident x-ray polarization $\lambda$ and
photon energy $\hbar \omega$ can be determined as the probability of
electronic transitions from initial core states with the total angular
momentum $j$ to final unoccupied Bloch states

\begin{eqnarray}
\mu_j^{\lambda} (\omega) &=& \sum_{m_j} \sum_{n \bf k} | \langle \Psi_{n \bf k} |
\Pi _{\lambda} | \Psi_{jm_j} \rangle |^2 \delta (E _{n \bf k} - E_{jm_j} -
\hbar \omega ) \nonumber \\
&&\times \theta (E _{n \bf k} - E_{F} ) \, ,
\label{mu}
\end{eqnarray}
where $\Psi _{jm_j}$ and $E _{jm_j}$ are the wave function and the
energy of a core state with the projection of the total angular
momentum $m_j$; $\Psi_{n\bf k}$ and $E _{n \bf k}$ are the wave
function and the energy of a valence state in the $n$-th band with the
wave vector {\bf k}; $E_{F}$ is the Fermi energy.

$\Pi _{\lambda}$ is the electron-photon interaction
operator in the dipole approximation
\begin{equation}
\Pi _{\lambda} = -e \mbox{\boldmath$\alpha $} \bf {a_{\lambda}},
\label{Pi}
\end{equation}
where $\bm{\alpha}$ are the Dirac matrices and $\bf {a_{\lambda}}$ is
the $\lambda$ polarization unit vector of the photon vector potential,
with $a_{\pm} = 1/\sqrt{2} (1, \pm i, 0)$,
$a_{\parallel}=(0,0,1)$. Here, $+$ and $-$ denotes, respectively, left
and right circular photon polarizations with respect to the
magnetization direction in the solid. Then, x-ray magnetic circular
and linear dichroism are given by $\mu_{+}-\mu_{-}$ and
$\mu_{\parallel}-(\mu_{+}+\mu_{-})/2$, respectively.  More detailed
expressions of the matrix elements in the electric dipole
approximation in the frame of the fully relativistic Dirac LMTO method may
be found in Ref.~\cite{book:AHY04}.

Concurrent with the development of x-ray magnetic circular
dichroism experiments, some important magneto-optical sum rules have
been derived \cite{LT88,TCS+92,CTA+93,LT96}.

For the $L_{2,3}$ edges the $l_z$ sum rule can be written as
\cite{book:AHY04}

\begin{equation}
\langle l_z\rangle = n_h \frac{4 \int_{L_3 + L_2} d \omega(\mu_+ - \mu_-)} {3
\int_{L_3 + L_2} d \omega(\mu_+ + \mu_-)} \,
\label{l_z}
\end{equation}
where $n_h$ is the number of holes in the $d$ band $n_h=10-n_{d}$, $\langle
l_z\rangle$ is the average of the magnetic quantum number of the orbital
angular momentum. The integration is taken over the whole 2$p$ absorption
region. The $s_z$ sum rule can be written as

\begin{eqnarray}
&&\langle s_z\rangle +\frac{7}{2}\langle t_z\rangle =
\label{s_z}
\nonumber \\  
&& n_h \frac{ \int_{L_3} d \omega(\mu_+ - \mu_-) -2 \int_{L_2} d
\omega(\mu_+ - \mu_-)} {\int_{L_3 + L_2} d \omega(\mu_+ + \mu_-)} \,
\end{eqnarray}
\noindent
where $t_z$ is the $z$ component of the magnetic dipole operator ${\bf
t}= {\bf s} - 3{\bf r}({\bf r}\cdot {\bf s})/|{\bf r}|^2$ which
accounts for the asphericity of the spin moment. The integration
$\int_{L_3}$ ($\int_{L_2}$) is taken only over the 2$p_{3/2}$
(2$p_{1/2}$) absorption region.

\subsection{Computational details}

The X-ray spectra~\cite{pardini2012cpc} are calculated using the
eigenvalues and wavefunctions of self-consistent electronic structures
calculations, which are carried out within DFT or DFT+$U$ using a
full-potential linear augmented plane wave (FLAPW)
method~\cite{wien2k} as well as the fully relativistic linear
muffin-tin orbital (RLMTO) method \cite{And75,PYLMTO}. This
implementation of the LMTO method uses four-component basis functions
constructed by solving the Dirac equation inside an atomic sphere
\cite{NKA+83}. The generalized gradient approximation (GGA)~\cite{perdew1996prl} was used for the
correlation and exchange potentials.  As for DFT+$U$, we use both
fully-localized-limit
(FLL)~\cite{liechtenstein1995prb,anisimov1993prb} and
around-the-mean-field (AMF)~\cite{czyzyk1994prb} double-counting
schemes with correlation parameters $U$ and $J$ applied on the
cation-$3d$ orbitals. We also used in this work the "relativistic"
generalization of the rotationally invariant version of the LSDA+$U$
method (RG) \cite{YAF03} which takes into account SO coupling so that
the occupation matrix of localized electrons becomes non-diagonal in
spin indices. 

In the x-ray absorption process an electron is promoted from a core
level to an unoccupied state, leaving a core hole. As a result, the
electronic structure at this state differs from that of the ground
state. In order to reproduce the experimental spectrum the
self-consistent calculations should be carried out including a core
hole. In this study the core-hole effect was fully taken into account
in the self-consistent iterations by removing an electron at the core
orbital using the supercell approximation. The core state of the
target atom in the ground state provides the initial state $|i>$ for
the spectral calculation. The final states $|f>$ are the conduction
band states obtained separately by the calculations in which one of
the core electrons of the target atom is placed at the lowest
conduction band. The interaction and the screening of the
electron-hole pair are fully accounted for by the self-consistent
iterations of the final state Kohn-Sham equations. This procedure
simulates the experimental situation, in which the sample can easily
supply an electron to screen a localized charge produced by the core
hole. Such an approach considers the symmetry breaking of the system
in a natural way, and self-consistently describes the charge
redistribution induced by the core hole. 

We employed the experimental lattice
parameters~\cite{mcguire2015cm,carteaux1995jpcm,deiseroth2006ejic} for
calculations. We used hexagonal cells which have six formula units,
with three layers instead of rhombohedral primitive cells for $\cri$
and $\cgt$.  For the thin film calculations, we took one unit cell of
hexagonal structures for $\cri$, $\cgt$ and an 1$\times$1$\times$2
supercell for $\fgt$ and added a 25 \textit{a.u.} vacuum region.  In
all of the bulk and film calculations, we did not perform any
structural optimization calculations.

\section{Results \& Discussion}
\subsection{$\fgt$}
\subsubsection {Electronic structure}

\begin{figure}[hbt]
	\centering
	\includegraphics[width=0.35\linewidth]{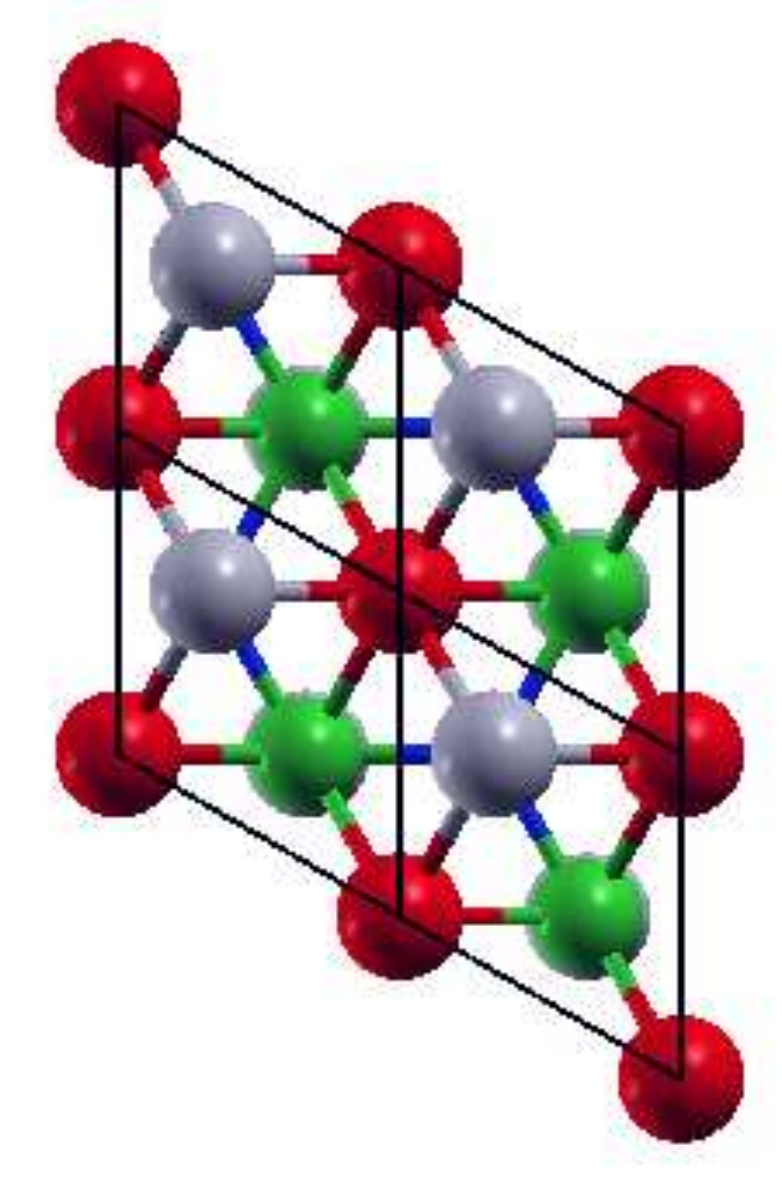}
	\includegraphics[width=0.35\linewidth]{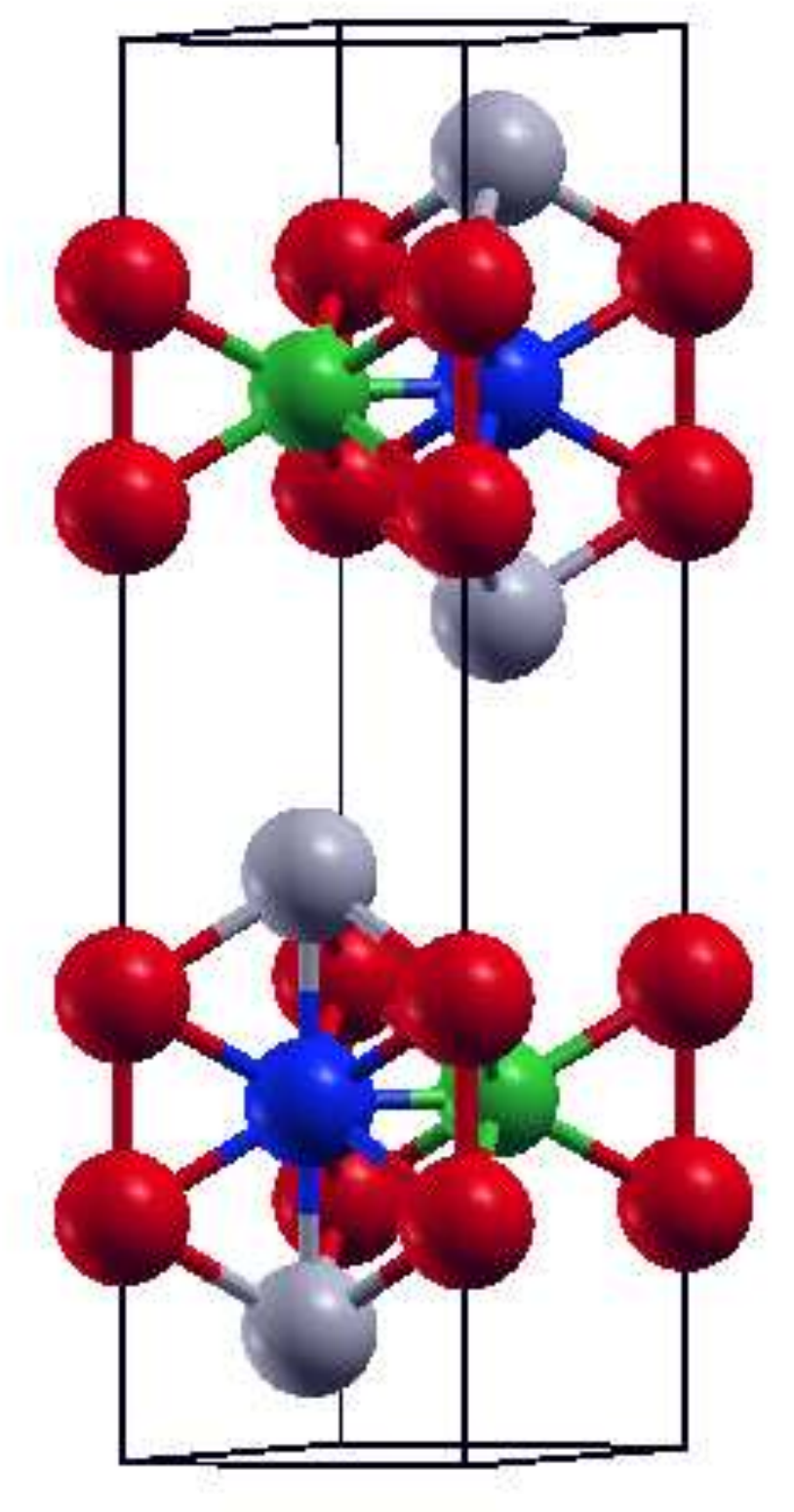}
	\caption{Schematic crystal structure of $\fgt$. The Fe$_{1}$
          atoms are red, Fe$_{2}$ atoms are blue, the Ge atoms are
          green and the Te atoms are gray. The top view (left) shows a
          2$\times$2$\times$1 supercell and the side view (right)
          shows a primitive unit cell. We used
          Xcrysden~\cite{kokalj1999176} for generating the structure
          figure.}
	\label{fig:fgt-struct}
\end{figure}

$\fgt$ crystallizes in a hexagonal ($P63/mmc$, space group no.~194)
structure~\cite{deiseroth2006ejic,may2016prb}.  The primitive cell
contains two formula units (f.u.).  As shown in \rfig{fig:fgt-struct},
Te atom occupies the $4f$($3m$) site and a Ge atom occupies the $2d$($-6m2$)
site, while Fe atoms are divided into two sublattices, $4e(3m)$ and
$2c(-6m2)$, denoted as Fe$_{1}$ and Fe$_{2}$, respectively.  Each
sublattice forms a trigonal lattice in the basal plane.  Fe$_{2}$ and
Ge atoms, together, form a Fe$_2$-Ge honeycomb monolayer; sandwiched between two Te-capped Fe$_{1}$ layers. 
The nearest neighbor of an Fe$_{1}$ atom is an Fe$_{1}$ atom along the $z$ direction and their distance is
\SI{2.554}{\AA}, and the distance between Fe$_{2}$ and its nearest
neighbor Ge is \SI{2.304}{\AA}.  

\begin{figure}[tbp!]
\begin{center}
\includegraphics[width=.9\columnwidth]{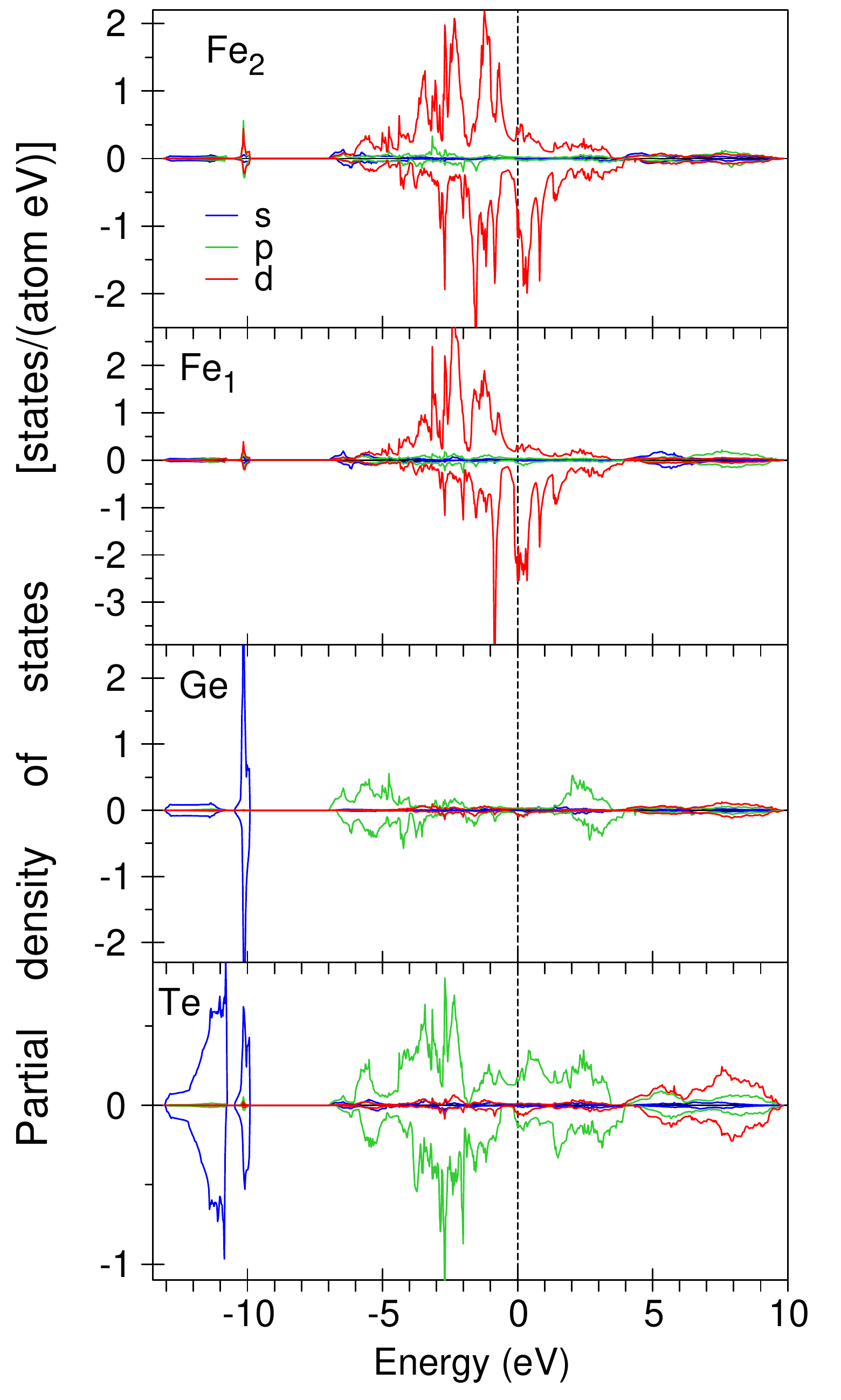}
\end{center}
\caption{\label{PDOS_FGT} The partial DOSs of
 bulk $\fgt$ calculated in the GGA approximation. Positive (Negative) DOS are up (down) spin states.}
\end{figure}

Figure \ref{PDOS_FGT} presents the partial density of states of $\fgt$
in the GGA approximation. The contribution of the Te 5$s$ states consists of two peaks that are located mostly between $-$13.1 and $-$10.8 eV
below the Fermi level. The second double peak located between $-$10.5
and $-$9.7 eV originates from the hybridization with Ge 4$s$
states. The Ge 4$s$ bands are located mostly between $-$10.5 and
$-$9.9 eV. The low intensity wide fine structure situated between
$-$13.1 and $-$10.8 eV is due the hybridization with Te 5$s$
states. The 5$p$ states of Te are found to be in the $-$7 eV to 3.5 eV
energy interval. The 4$p$ states of Ge occupy the same
energy interval, although, they have relatively small intensity in the
vicinity of the Fermi level. The spin splittings of the Te and Ge $p$
states are quite small. The Te 5$d$ states are situated from 3.9 eV to
9.8 eV above the Fermi level. Although the Fe$_1$ and Fe$_2$ 3$d$
states situated between $-$6.6 eV and 4 eV energy interval are very
similar, they still have some differences. There is a strong spin-down
peak at $-$0.9 eV at the Fe$_1$ site, while the corresponding peak at the
Fe$_2$ site is much weaker. Besides, the Fe$_1$ site possesses a quite
intense spin-down peak in just above the Fermi level, while a similar
peak is much smaller at the Fe$_2$ site. Both the sites have quite
large spin-down 3$d$ density of states and small spin-up ones above
the Fermi level.

\begin{table}[hbt]
  \caption{On-site spin $\mspin$ and orbital $\morb$ magnetic moment
    of two different Fe sites in $\fgt$.  The calculations are
    performed with the GGA approximation using the FLAPW and RLMTO
    methods, as well as the AMF and FLL schemes of PBE+$U$.  }
  \label{tbl:fgt-bulk-mm}
\label{tbl:mi}
\bgroup
\def\arraystretch{1.1}
\begin{tabular*}{\linewidth}{l @{\extracolsep{\fill}} crrrrcccc}
  \hline
  \hline
  \\[-1em]
  {Bulk} & &$U$ &$J$ & &  \multicolumn{2}{c}{Fe$_1$}  &    &  \multicolumn{2}{c}{Fe$_2$} \\
  \\[-1em]  
  \cline{1-1} \cline{3-4} \cline{6-7} \cline{9-10}
  \\[-1.1em]
  Method   & &  (Ry) & (Ry) &  & $\mspin$   & $\morb$    &     & $\mspin$ & $\morb$ \\  
\\[-1.1em]
  \hline                                      
\\[-1.1em]
FLAPW & &        &  &  & 2.36 & 0.07 & & 1.55 & 0.03  \\
RLMTO & &        &  &  & 1.91 & 0.12 & & 1.26 & 0.02  \\
sum rules & &    &  &  & 2.11 & 0.11 & & 1.32 & 0.02  \\
AMF   & & 0.1    &  &  & 2.32 & 0.13 & & 1.59 & 0.05  \\
      & & 0.2    &  &  & 1.77 & 0.14 & & 1.65 & 0.08  \\
FLL   & & 0.1    &  &  & 2.62 & 0.08 & & 1.81 & 0.07  \\
      & & 0.2    &  &  & 2.82 & 0.11 & & 2.09 & 0.13  \\
Exp\footnotemark[1] & &         &  & \multicolumn{2}{c}{2.18}  & & \multicolumn{2}{c}{1.54}  \\
\\[-1.0em]
  \hline
  {Film} & &    & &  &  &    & & \\
  \hline  
		FLAPW & &    &     &  & 2.35 & 0.07 & & 1.55 & 0.03  \\
		FLL & & 0.2 &0.00    &  & 2.80 & 0.11 & & 2.08 & 0.13  \\
	    	& & 0.2 &0.10    &  & 2.60 & 0.12 & & 1.84 & 0.12  \\
            & & 0.2 &0.15    &  & 2.48 & 0.17 & & 1.71 & 0.12  \\
		\\[-1.0em]
		\hline\hline
 
\end{tabular*}
\egroup \footnotetext[1]{Total magnetic moment measured using neutron
  powder diffraction by May $\etal$~\cite{may2016prb}.}
\end{table}

The on-site spin and orbital magnetic moments of both Fe sites in bulk
and thin-film $\fgt$, calculated using various methods, are listed in
\rtbl{tbl:fgt-bulk-mm} and compared with the on-site total magnetic
moments from neutron powder diffraction.  Fe$_{1}$ has larger spin and
orbital moments than Fe$_{2}$ does, which is consistent with published
experimental and theoretical
results~\cite{may2016prb,verchenko2015inorgchem}.  Within the GGA
approximation, the calculated ratio between Fe$_{1}$ and Fe$_{2}$
total magnetic moments is equal to 1.54, which is larger than the
experimental value of 1.45~\cite{may2016prb} by about 10\%. GGA
underestimates the orbital moment by $\sim40\%$, in comparison with
the experimental value of $\morb$ = 0.10 $\mub$ obtained using the
XMCD~\cite{zhu2016prb}. However, fully relativistic LMTO method
produces an orbital magnetic moment in better agreement with the
experiment (Table \ref{tbl:mi}).

We also also present in the \rtbl{tbl:fgt-bulk-mm} the Fe spin and
orbital magnetic moments obtained by the sum rules [Eqs. (\ref{l_z})
  and (\ref{s_z})] applyed to the theoretically calculated XAS and
XMCD spectra in the GGA approximation in a frame of the RLMTO
method. Although the XMCD sum rules are derived within an ionic model
using a number of approximations and the application of the sum rules
sometimes results in an error up to 50\% \cite{book:AHY04}, we found
relatively good agreement between the theoretically calculated
magnetic moments and those derived from the sum rules.

To investigate the correlation effects on magnetism, we perform
GGA+$U$ calculations and checked if the on-site Coulomb interaction term
$U$ is able to improve agreement with experiments.  The results depend on the double counting
schemes used.  \rtbl{tbl:fgt-bulk-mm} shows calculated magnetic
moments with $U=0.1$ and \SI{0.2}{Ry} using both the AMF and FLL
schemes. The orbital moments ($\morb$) in both schemes are increased with
increasing $U$ parameter, improving the agreement with the experiment. While, for the
spin moments, the FLL scheme increases $\mspin$ on both sites, deviating further
from experiments, the AMF scheme decreases (increases) $\mspin$ on Fe$_{1}$ (Fe$_{2}$) sites giving better agreement with the experiment.  
Thus, overall, a small $U$ parameter within the AMF scheme
improves the agreement with the experiment.

We also investigate the dependence of Hunds's $J$ parameter and the
surface effects.  The magnetic moments of the bulk and film are nearly
identical, reflecting the van der Waals nature of the system.
\rtbl{tbl:fgt-bulk-mm} also shows the dependence of the magnetic
moments on the Hunds's $J$ parameter in film $\fgt$.  With increasing
of $J$ values, spin moments are decreased at both iron sites while the
orbital moment is increased only on the Fe$_{1}$ site. Thus the spin
fluctuation introduced by Hund's $J$ improves the agreement with
experiments.  
Our work suggests that using a single value for the $U$ parameter may not be sufficient for describing this system. An explicit treatment of electron correlations beyond DFT for these systems may be valuable~\cite{lee2020prb,ke2020spinexcitation}.

\subsubsection{The XAS and XMCD spectra}

\begin{figure}[hbt]
  	\centering
    \includegraphics[width=.9\linewidth]{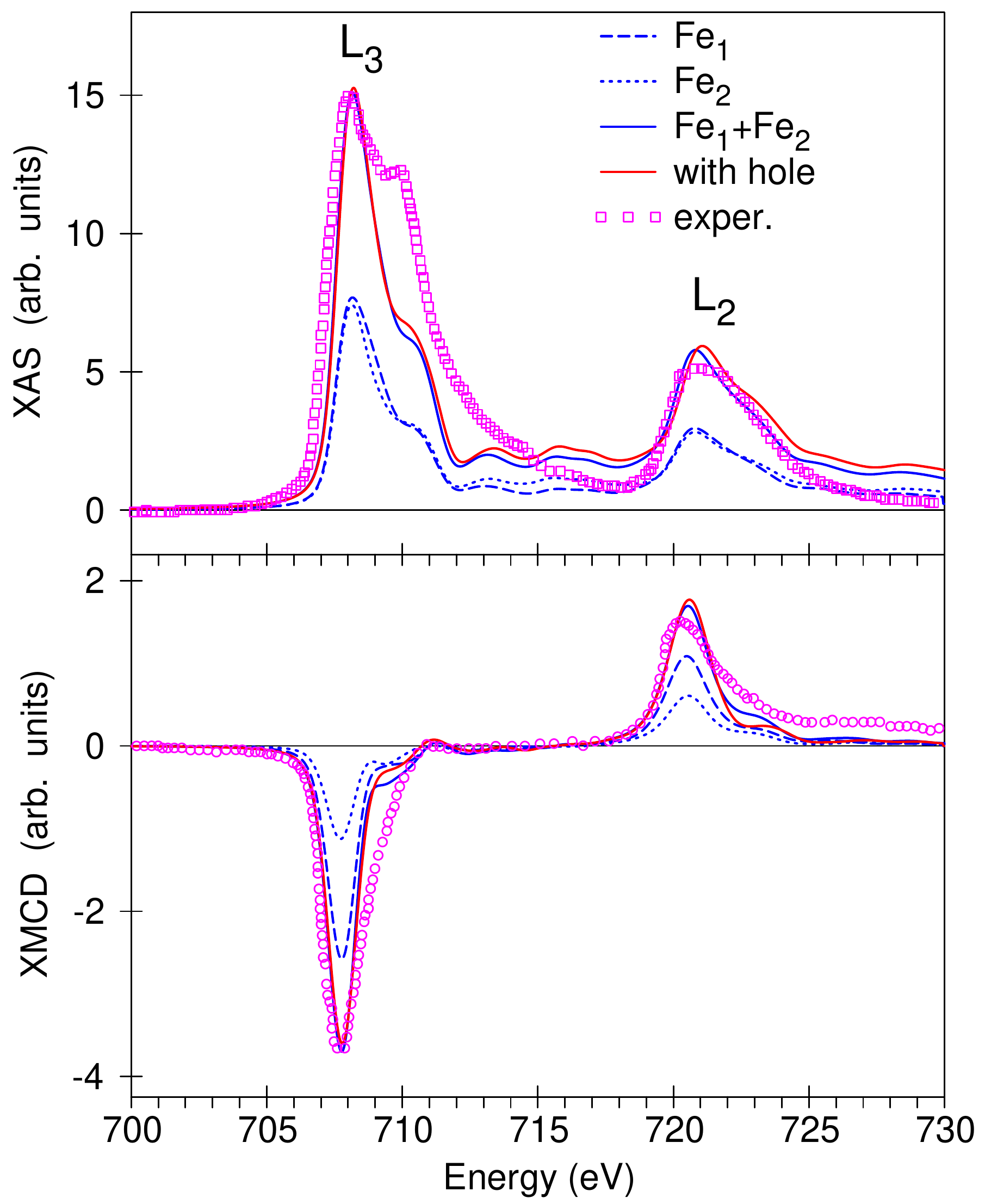}
  	\caption{Comparison calculated Fe $L_{2,3}$-edges XAS and XMCD
          spectra of bulk $\fgt$ with experiment. The solid red (blue) line is with (without) core-hole effects. The experimental data are obtained from Ref.~[\onlinecite{zhu2016prb}].}
  	\label{Fe_L23}	
  \end{figure}

Figure \ref{Fe_L23} presents the calculated XAS as well as XMCD
spectra of the $\fgt$ compound at the Fe $L_{2,3}$ edges in the GGA
approximation compared with the experimental data
\cite{zhu2016prb}. The XMCD spectra at the Fe $L_{2,3}$ edges are
mostly determined by the strength of the spin-orbit (SO) coupling of
the initial Fe 2$p$ core states and spin-polarization of the final
empty 3$d_{3/2,5/2}$ states while the exchange splitting of the Fe
2$p$ core states as well as the SO coupling of the 3$d$ valence states
are of minor importance for the XMCD at the Fe $L_{2,3}$ edges of
$\fgt$. Because of the dipole selection rules, apart from the
4$s_{1/2}$ states (which have a small contribution to the XAS due to
relatively small 2$p$ $\to$ 4$s$ matrix elements) only 3$d_{3/2}$
states occur as final states for $L_2$ XAS for unpolarized radiation,
whereas for the $L_3$ XAS the 3$d_{5/2}$ states also
contribute. \cite{book:AHY04} Although the 2$p_{3/2}$ $\to$ 3$d_{3/2}$
radial matrix elements are only slightly smaller than for the
2$p_{3/2}$ $\to$ 3$d_{5/2}$ transitions the angular matrix elements
strongly suppress the 2$p_{3/2}$ $\to$ 3$d_{3/2}$
contribution \cite{book:AHY04}. Therefore neglecting the energy
dependence of the radial matrix elements, the $L_2$ and the $L_3$
spectrum can be viewed as a direct mapping of the DOS curve for
3$d_{3/2}$ and 3$d_{5/2}$ character, respectively.

The experimental Fe $L_3$ XAS has one prominent peak around 708 eV and
a pronounced shoulder bump at around 710 eV shifted by about 2 eV with
respect to the maximum to higher photon energy. This structure is less
pronounced at the $L_2$ edge. This result can be ascribed to the
lifetime broadening effect because the lifetime of the 2$p_{1/2}$ core
hole is shorter than the 2$p_{3/2}$ core hole due to the $L_2L_3V$
Coster-Kronig decay. The GGA approximation reasonably well describes
the shape of the XAS spectra at the Fe $L_{2,3}$ edges (the upper
panel of Fig.\ \ref{Fe_L23}), however it underestimates the high
energy peak at around 710 eV. The calculated spectra have also smaller
width compared to the experimental spectra.

The lower panel of the Fig. \ref{Fe_L23} shows XMCD spectra of the
$\fgt$ compound at the Fe $L_{2,3}$ edges in GGA approximation
compared with the experimental data \cite{zhu2016prb}. The Fe$_{1}$
site shows stronger XMCD spectra than the Fe$_{2}$ site due to
larger orbital magnetic moment at the Fe$_1$ site in comparison with
the Fe$_2$ one.

We found minor influence of the final-state interaction on the shape
of the Fe $L_{2,3}$ XAS and XMCD spectra in the whole energy
interval (red curves in Fig. \ref{Fe_L23}). 

\begin{figure}[hbt]
	\centering
	\includegraphics[width=0.9\linewidth]{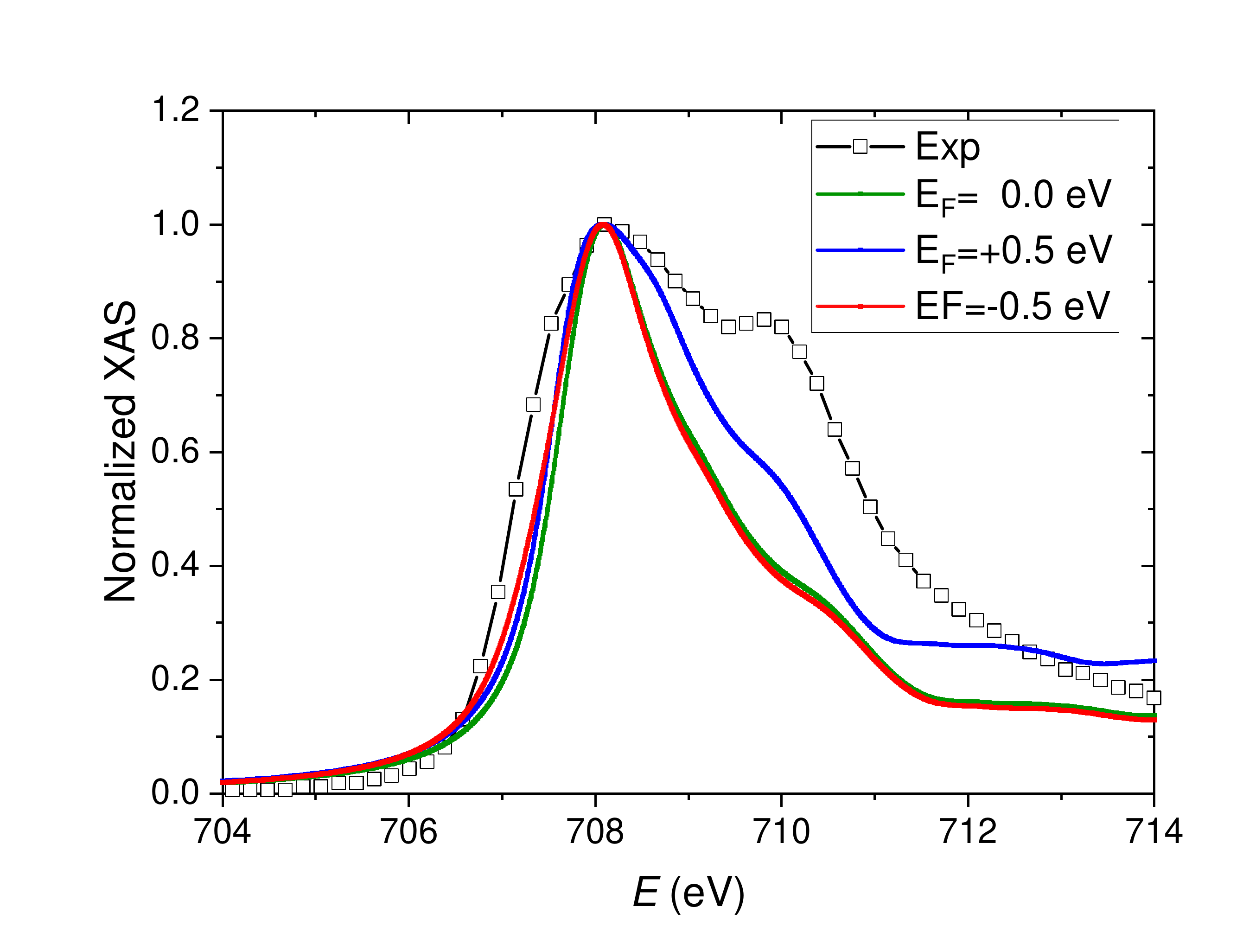}
	\caption{Comparison of bulk XAS of $\fgt$ with shifting
          E$_{F}$. The green is the spectral line shape with PBE and
          the blue (red) is the result that E$_{F}$ is shifted 0.5eV
          up (down). The line with open squares is the experimental
          result~\cite{zhu2016prb}. }
	\label{fig:efchange}	
\end{figure}

Zhu $\etal$~\cite{zhu2016prb} compared total DOSs calculated by DFT and DFT+DMFT.  Their DFT result shows a broad peak right above E$_{F}$ and two sharp peaks at higher energies. However, in the DFT+DMFT calculation, the E$_{F}$ is located at the peak which is much narrower than the DFT peak near
E$_{F}$. It is possible that DFT overcounts down spin empty states just above E$_{F}$ so that it overemphasizes the first peak of XAS. To check this scenario, we calculated the spectra with shifted E$_{F}$.  Shifting E$_{F}$ can mimic the
change in the number of electrons and mimic doping. It is interesting
to find the change of relative intensity between the two peaks with
E$_{F}$ changes.  \rFig{fig:efchange} shows XAS with the GGA
functional and with $\pm{0.5}$ $eV$ shifted E$_{F}$ from GGA results.
As expected, shifting E$_{F}$ up reduces unoccupied states of down
spin and the intensity of the first peak of the spectra.  It results
in the second peak becoming relatively stronger than for the GGA result.

\begin{figure}[hbt]
	\centering
	\includegraphics[width=0.9\linewidth]{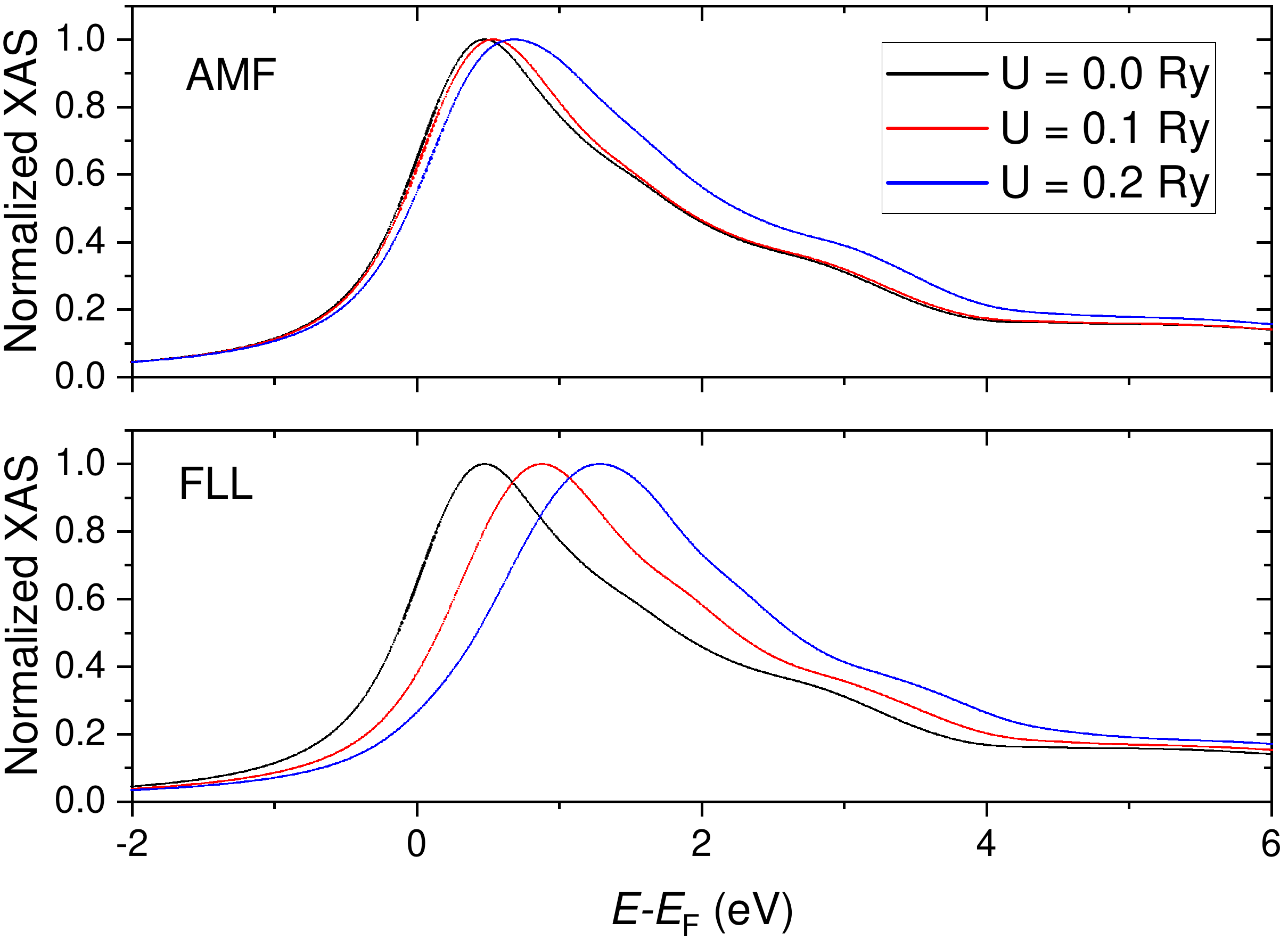}
	\caption{Calculated $\fgt$ bulk XAS spectra with different $U$
          values. The top (bottom) panel shows results of the AMF
          (FLL) scheme of GGA+$U$ method. }
	\label{fig:fgt-bulk-ldau}	
\end{figure}

Since the XAS spectral line shape depends on details of DOS or band
structure and an on-site Coulomb interaction changes electronic
structure, it is interesting to find how the spectra change with
different $U$ values. We calculated the spectra with the around mean
field (AMF) and the fully localized limit (FLL) schemes of LDA+$U$ to
see how the spectra change with $U$ and $J$ values. For the AMF
scheme, the higher $U$ value gives broader XAS spectra and a
relatively higher second bump which is related to the lower spin
moment. The intensity of XMCD spectra also depends on $U$ values that
are attributed to moment change.  The top panel of
~\rfig{fig:fgt-bulk-ldau} shows results with the AMF scheme. As the
figure shows, higher $U$ values give broader XAS spectra shape and a
relatively higher second bump which is related to the lower spin
moment (See~\rtbl{tbl:fgt-bulk-mm}). The bottom panel of
~\rfig{fig:fgt-bulk-ldau} shows results of the FLL scheme with $U=0.1$
and \SI{0.2}{Ry} but $J=\SI{0}{\eV}$. For XAS spectra, instead of
getting a broader line shape, the main peaks are moving toward higher
energy. With increasing $J$ value, XAS is slightly shifted toward
higher energy but XMCD becomes narrower (not shown).  Although there
are changes in detail of the spectra, overall spectral line shapes
change little. The GGA+$U$ method produces altered electronic
structure compared to the GGA functional but the effects of $U$ on the XAS
(XMCD) spectra line shape are not significant.

\subsection{$\cri$}
\subsubsection {The electronic structure}

\begin{figure}[tbp!]
	\begin{center}
		\includegraphics[width=.9\columnwidth]{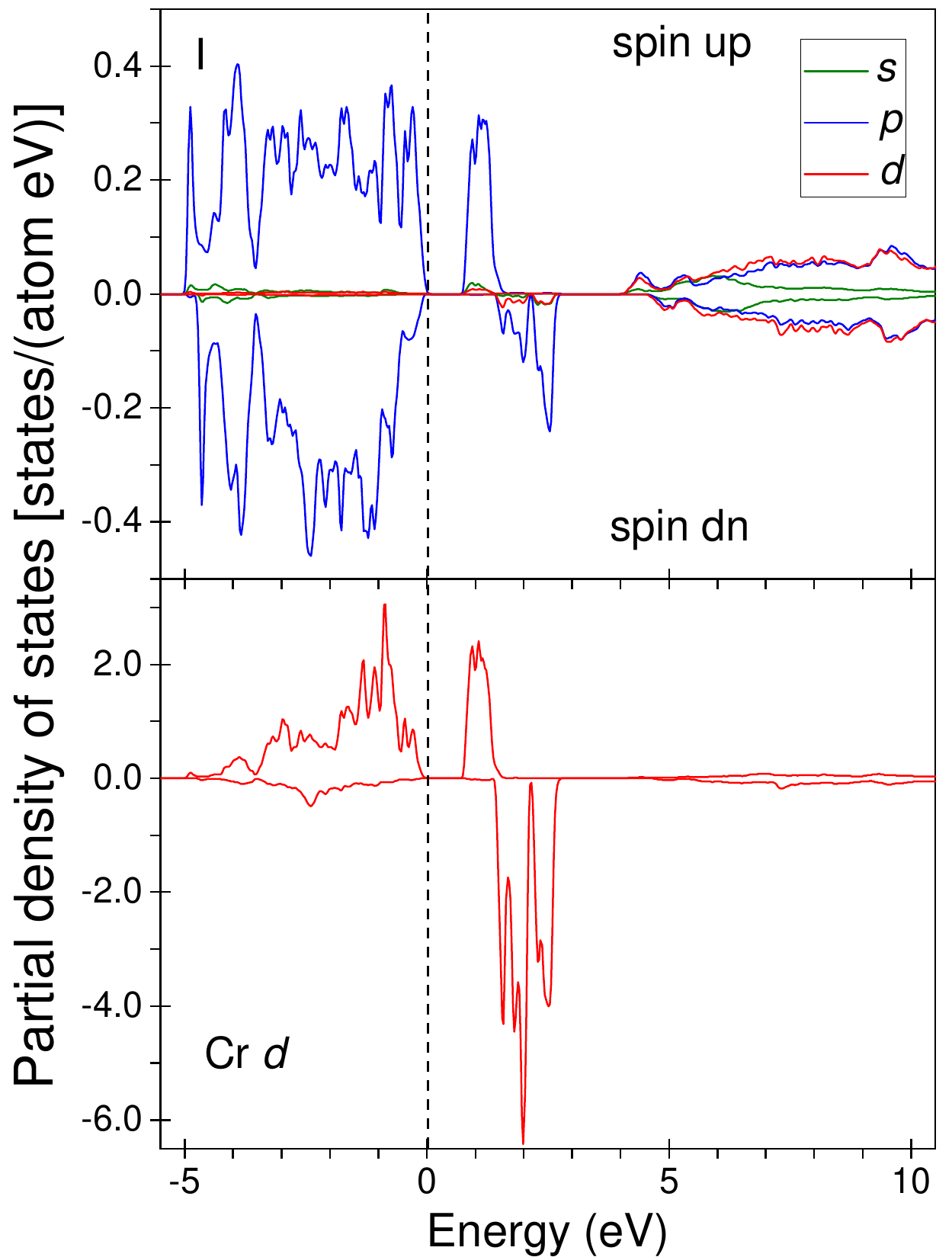}
	\end{center}
	\caption{\label{PDOS_CrI3} The partial DOSs of
		bulk $\cri$ calculated in the GGA approximation. }
\end{figure}

Figure \ref{PDOS_CrI3} presents the partial density of states of
$\cri$ in the GGA approximation. 
The 5$p$ states of I are located in the $-$5.1 eV to 2.7 eV energy
interval. It is interesting to note that the 5$s$ partial DOS of I is extremely small and the number of the I 5$s$ electrons is equal to 0.02 in $\cri$ instead of 2 in the free I atom. The Cr 3$d$ spin up states are situated in the  $-$5.1 eV and 1.5 eV energy interval. The empty Cr 3$d$ spin-up DOS shows up as a single peak in the 0.7eV to 1.5 eV interval. The empty Cr 3$d$ spin-down DOSs consist of two
narrow intensive peaks in the 1.4 eV to 2.1 eV and 2.2 eV to 2.6
intervals, respectively.

\rtbl{tbl:cri3-bulk-mm} and \rtbl{tbl:cri3-bulk-mm-sic} show
calculated Cr magnetic moments and band gaps of bulk $\cri$ with
various $U$ and $J$-parameters. While the AMF scheme gives lower spin
moment with increasing $U$ parameters, the FLL scheme produces changes
in the opposite direction. Gudelli $\etal$~\cite{gudelli2019njp}
performed GGA+$U$ calculations for bulk, mono-,bi-, and tri-layer of
$\cri$ and found the orbital moments are parallel to spin moments for
the Cr atom. This agrees with the result that is presented
here. However, the results that are estimated from XMCD spectra and
atomic calculations ~\cite{frisk2018materialletters,kim2019prl} are in
the opposite direction compared to GGA and GGA+$U$ results.
The magnetic moment of Cr in a thin film is not so different from the
bulk case (See~\stbl{1} in SI). It is because the Cr atom layers are
shielded by I atom layers. If Cr atoms are located on the top surface,
they may acquire a larger moment but because of Van der Waals bonding
character - the inter-layer interaction is much weaker than the intra-layer interaction, it is not plausible to obtain structures where Cr atoms are on the top surface.
 
\begin{table}[hbt]
	\caption{On-site spin $\mspin$ and orbital $\morb$ magnetic
          moment of Cr of $\cri$. FLL is calculated with
          $J=\SI{0}{Ry}$. Experimental orbital moments are obtained by
          XMCD measurement.}
	\label{tbl:cri3-bulk-mm}
	\bgroup
	\def\arraystretch{1.1}
	\begin{tabular*}{\linewidth}{l @{\extracolsep{\fill}} crrrrccc}
		\hline \hline \\[-1em] {$\cri$} & & $U$ & &
                \multicolumn{2}{c}{Cr} & & gap \\ \\[-1em] \cline{1-1}
                \cline{3-3} \cline{5-6} \cline{8-8} \\[-1.1em] Method
                & & Ry & & $\mspin$ & $\morb$ & & eV \\ \\[-1.1em]
                \hline \\[-1.1em]
                FLAPW & & & & 2.99 & 0.074 & & 0.77\\
                RLMTO & & & & 3.23 & 0.118 & & 0.42 \\
            sum rules & & & & 2.74 & 0.108 & &      \\
                AMF & & 0.1 & & 2.84 & 0.078 & & 1.06 \\
                & & 0.2 &   & 2.63 & 0.080 & & 1.31 \\
		    & & 0.4     &  & 2.12 & 0.074 & & 0.72   \\
		FLL & & 0.1     &  & 3.17 & 0.070 & & 0.74   \\
		    & & 0.2     &  & 3.22 & 0.068 & & 0.72   \\
		    & & 0.4     &  & 3.42 & 0.062 & & 0.65   \\
		Exp~\cite{kim2019prl} &&    &&  &-0.059&&  \\
		\\[-1.0em]
		\hline\hline
	\end{tabular*}
	\egroup
\end{table}

\begin{table}[hbt]
	\caption{$U$ and $J$ dependency of on-site spin $\mspin$ and orbital $\morb$ magnetic moment of Cr of $\cri$}
	\label{tbl:cri3-bulk-mm-sic}
	\bgroup
	\def\arraystretch{1.1}
	\begin{tabular*}{\linewidth}{l @{\extracolsep{\fill}} crrrrccc}
		\hline
		\hline
		\\[-1em]
		{$\cri$} & & $J$ & &  \multicolumn{2}{c}{Cr}  &    &  gap \\
		\\[-1em]  
		\cline{1-1} \cline{3-3} \cline{5-6} \cline{8-8}
		\\[-1.1em]
		$U$ (Ry)   & & Ry &  & $\mspin$   & $\morb$    &     & eV \\
		\\[-1.1em]
		\hline                                      
		\\[-1.1em]
	   0.30 & & 0.05    &  & 3.14 & 0.066 & & 0.90   \\
		    & & 0.10    &  & 2.85 & 0.065 & & 1.30   \\
		    & & 0.20    &  & 2.15 & 0.055 & & 0.68   \\
       0.40 & & 0.05    &  & 3.17 & 0.062 & & 0.86   \\
            & & 0.10    &  & 3.22 & 0.061 & & 1.29   \\
            & & 0.20    &  & 3.42 & 0.062 & & 0.60   \\
		\\[-1.0em]
		\hline\hline
	\end{tabular*}
	\egroup
\end{table}

\subsubsection {The XAS and XMCD spectra}
\begin{figure}[hbt]
	\centering	
	\includegraphics[width=1.0\linewidth]{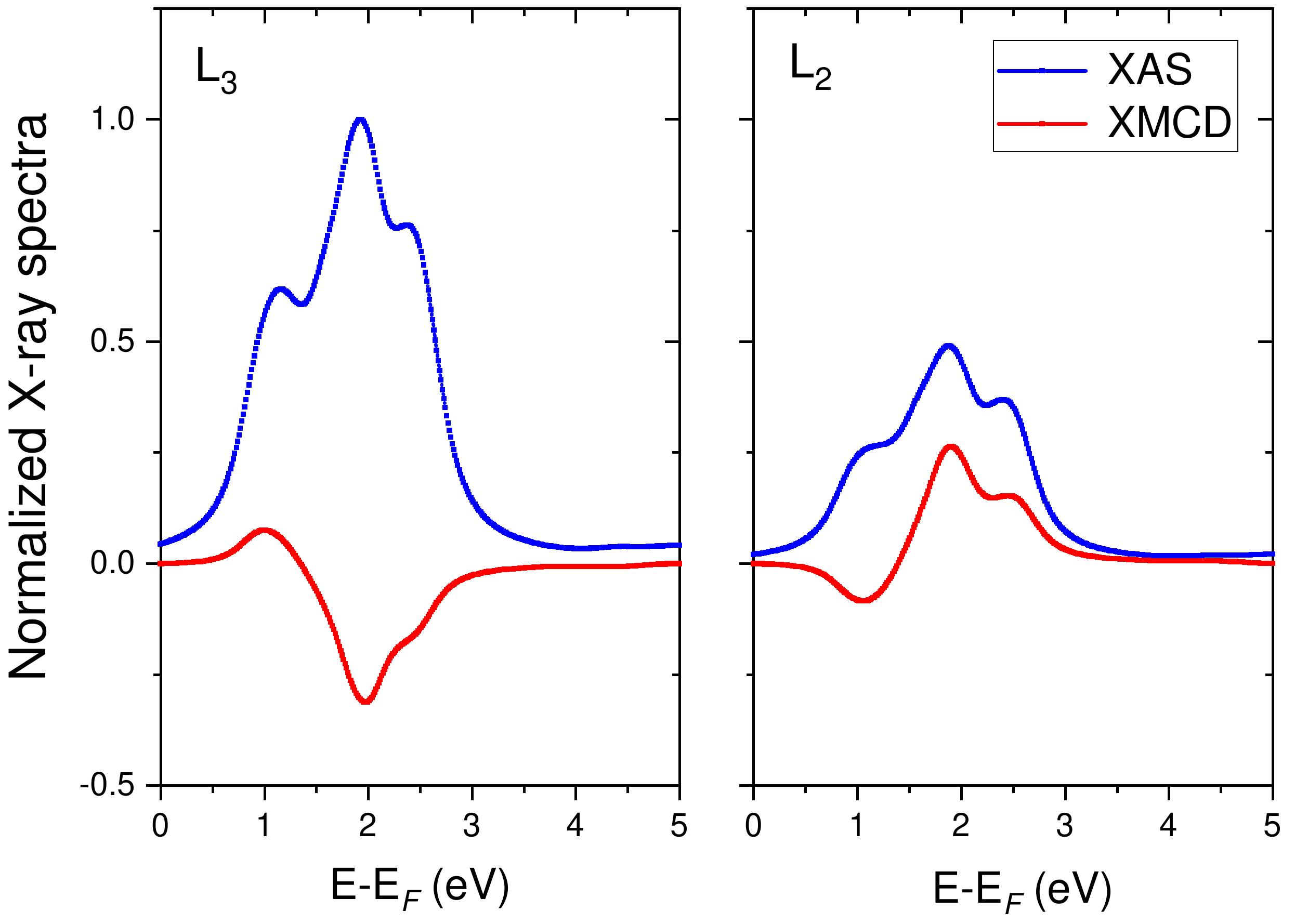}
	\caption{ Calculated L$_{3}$ (left panel) and L$_{2}$ (right
          panel) edge XAS (blue) and XMCD (red) spectra in bulk
          CrI$_{3}$. The PBE functional was employed for the
          calculation. Each spectra shows three structures.}
	\label{fig:spectra-cri3-bulk}	
\end{figure}

\rFig{fig:spectra-cri3-bulk} shows calculated XAS and XMCD spectra
with GGA functionals. The left panel is L$_{3}$-edge spectra and the
right panel is L$_{2}$-edge spectra. Both XASs show three structures~$\colon$a peak around 1 eV, a central peak around 1.9 eV and a
shoulder bump around 2.4 eV. XMCD spectra also have three features -
the first has a different sign compared to the other two peaks.

Frisk $\etal$'s~\cite{frisk2018materialletters} measured L$_{3}$ edge
XAS shows a strong peak around 576 eV photon energy and two bumps on
both sides of the peak.  However, their calculated spectra does not
show the bump at higher energy but a smooth decreasing from the central
peak. With these results, they suggested that the higher energy bump
is attributed to partial oxidation.  Kim $\etal$'s~\cite{kim2019prl}
results look similar to Frisk $\etal$'s except for the bump at higher
energy. 
They observed a small bump and were able to reproduce it by their model calculation.
It seems that the third peak is intrinsic and small, although it can be intensified by oxidation.

\begin{figure}[hbt]
	\centering	
	\includegraphics[width=1.0\linewidth]{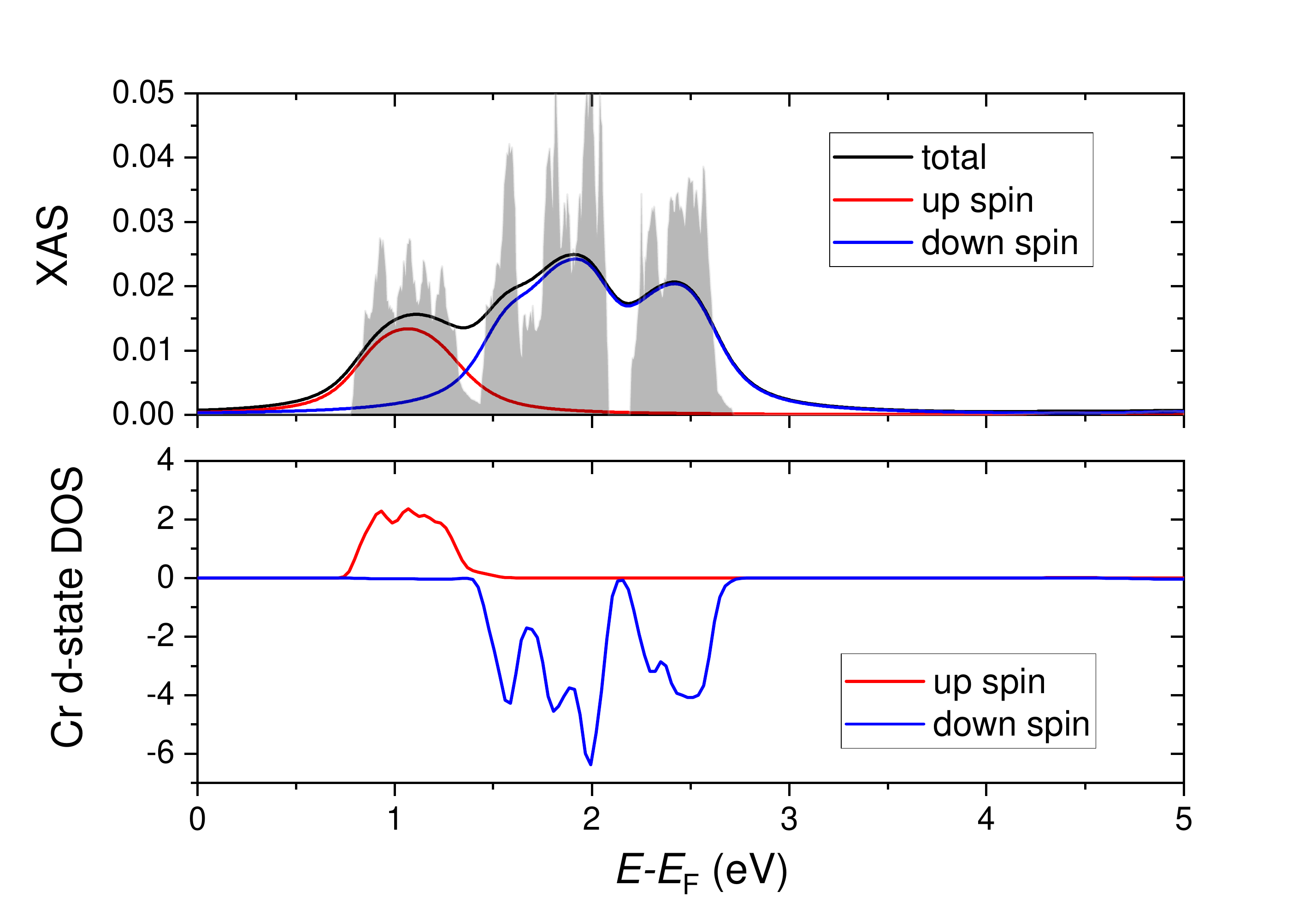}
	\caption{The top panel shows CrI$_{3}$ bulk L$_{3}$-edge XAS
          spectra calculated by 
 $\mu_{xas}= M^{\uparrow 2}(E)\rho^{\uparrow}(E)+ M^{\downarrow 2}(E)\rho^{\downarrow}(E)$.
 where $M$and $\rho$ is radial matrix elements and density of state respectively.
 Red (blue) line  presents up (down) spin contribution and shadow region show without broadening effects.  The bottom panel shows spin-decomposed Cr $d$-state DOS of $\cri$.  See~\sfig{2} in SI for the spin dependent radial matrix elements of $\cri$.}
	\label{fig:spectra-cri3-bulk-dos}	
\end{figure}

The top panel of ~\rfig{fig:spectra-cri3-bulk-dos} shows Cr
L$_{3}$-edge XAS which was calculated using radial matrix elements and
DOS.  It not only reproduces the three features of
~\rfig{fig:spectra-cri3-bulk} well but also gives more information on
the character of the peaks if it is combined with DOS (the bottom
panel of ~\rfig{fig:spectra-cri3-bulk-dos}).  The spectra were
decomposed into up spin (red) and down spin (blue) contributions.  It
shows the first peak is attributed to the up spin state and the other
two peaks are from the down spin state. It also explains why the first
peak of the XMCD spectra has the opposite sign of the other two
peaks.


The main difference between the measured XAS and the calculated XAS
with GGA 
is the spectral width. The
calculated spectra has a much narrower spectral width than the
experimental result. The spectral intensity of the calculated spectra
is rapidly decreased after the third bump.  It is because Cr
3\textit{d} bandwidth is narrow and the top of the band is located
about 2.5 eV from E$_{F}$ which corresponds to about 576.6 eV.
The sign of the calculated XMCD peaks is also not consistent with the
experiment.  While the experimental result shows a negative sign for
the first and second peaks and a positive sign for the third peak, the
calculated spectra show a positive sign for the first peak and
negative for the others.

The XMCD spectra depends on details of \textit{m$_{l}$} decomposed DOS.
The GGA+$U$ method is able to
adjust relative positions of \textit{m$_{l}$} decomposed DOS by
controlling $U$ values. Therefore it is possible to tune
the calculated spectra by using different $U$ values.  We have
performed the GGA+$U$ calculations to understand the effects of $U$
and $J$ values on X-ray spectra and to check if the GGA+$U$ method is
able to produce X-ray spectra that is consistent with the experimental
results. 


\begin{figure}[hbt]
	\centering	
	\includegraphics[width=0.95\linewidth]{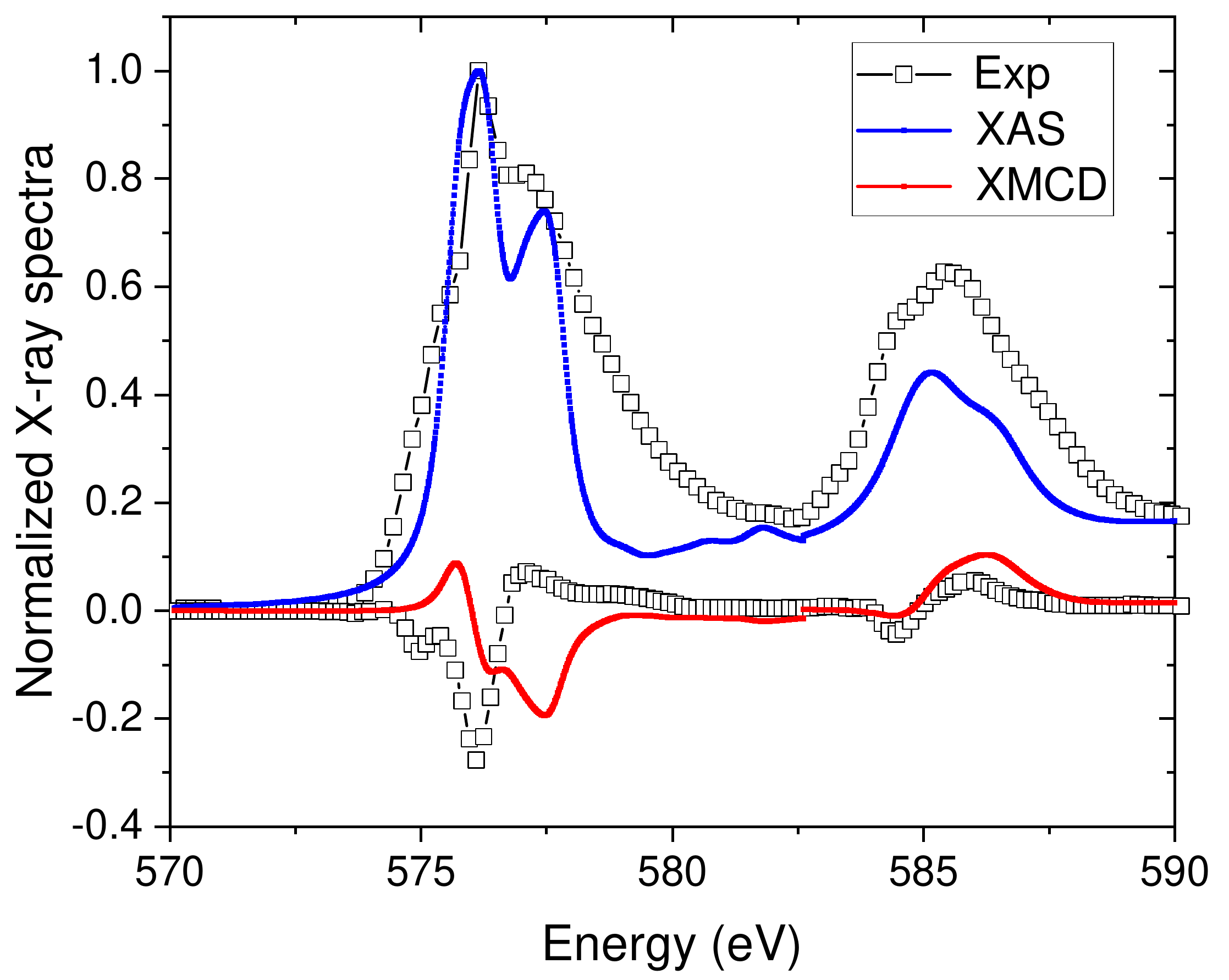}\\
	\includegraphics[width=0.95\linewidth]{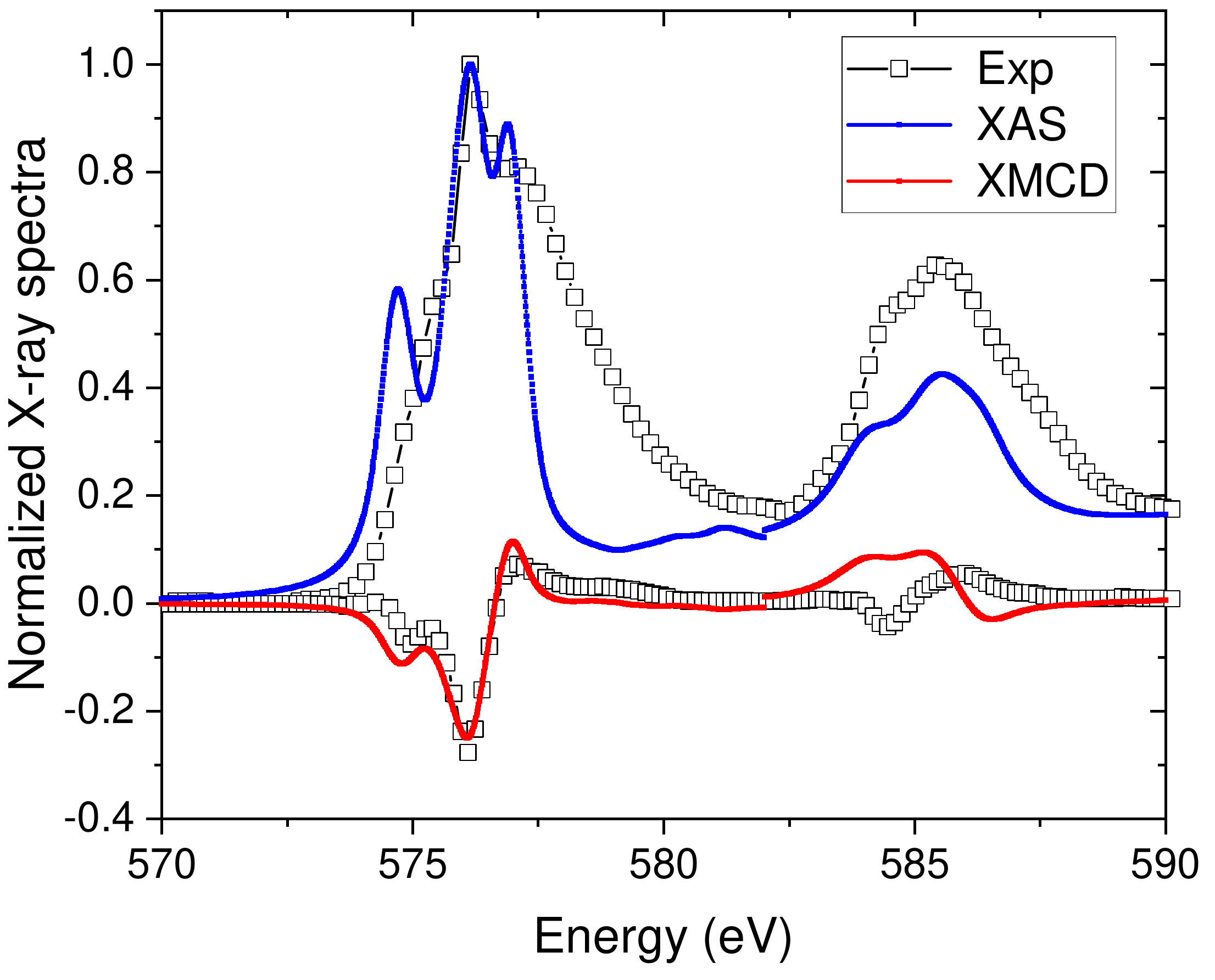}
	\caption{Calculated $\cri$ bulk L$_{3}$ and L$_{2}$-edge XAS
          (blue lines)and XMCD (red lines) spectra. The top panel
          shows results of FLL scheme with $U=\SI{0.4}{Ry}$ and
          $J=\SI{0.1}{Ry}$. The bottom panel shows AMF scheme results
          with $U=\SI{0.4}{Ry}$. The lines with open squares are
          experimental data which were obtained from Frisk
          $\etal$~\cite{frisk2018materialletters}. }
	\label{fig:cri3-bulk-xmcd-ldau1}
\end{figure}

\rFig{fig:cri3-bulk-xmcd-ldau1} shows the result with $U=\SI{0.4}{Ry}$ (See ~\sfig{3} for a lower $U$ value result).  For a comparison, it also includes
experimental data that are obtained from Frisk
$\etal$~\cite{frisk2018materialletters}.  Since higher $U$ parameters
promote 3$d$ band splittings to be stronger and band widths to be wider, the calculated spectra have wider spectral widths.  For the XAS spectra,
the AMF scheme gives better agreement with experiment than the FLL
scheme.  The calculation with the AMF scheme gives good agreement with
the experiment for the L$_{3}$-edge but not for the L$_{2}$-edge XMCD
spectra.  It is reversed from the calculation with the FLL scheme.
The calculated XMCD spectra with the FLL scheme shows reasonable
agreement for the L$_{2}$-edge but not for the L$_{3}$-edge.  

The normalized X-ray spectra of the thin film is similar to the case of
the bulk $\cri$ except for details of fine structure of L$_{3}$ XAS spectra.
See~\sfig{4} in SI for the film X-ray spectra.  The FLL scheme
with $U=\SI{0.4}{Ry}$ and $J=\SI{0.1}{Ry}$ is able to reproduce most
of the experimental features except for the higher energy peak (around
576.6 eV photon energy) of L$_{3}$-edge XMCD spectra. While the AMF
scheme with $U=\SI{0.4}{Ry}$ produces the L$_{3}$-edge XMCD spectra
which shows good agreement with the experimental result, it flips the
sign of lower energy peak of the L$_{2}$-edge XMCD spectra.  For the
L$_{3}$ edge XMCD spectra, the lowest energy peak which has a positive
sign is attributed to the up spin state and the other two peaks which
have negative signs are attributed to the down spin state.  The
positive peak is located at the highest energy in the L$_{3}$ edge
XMCD spectra calculated by the AMF scheme.
	
Overall, the GGA+$U$ method is able to improve theoretical spectra of the bulk $\cri$ but it requires a rather higher $U$ value.  
The results with $U\approx\SI{5.2}{eV}$ show good agreement with the measured spectra but this $U$ is much higher than the values from
published work.
For instance, the employed $U$ values are 1.0 eV in
Gudelli $\etal$~\cite{gudelli2019njp}, 2.0, 2.9 eV in Jang $\etal$'s
~\cite{jang2019prm}, and 3.0 eV in Sivadas $\etal$~\cite{sivadas2018nanoletter}.

\subsection{$\cgt$}
\subsubsection {The electronic structure}
\begin{figure}[tbp!]
\begin{center}
\includegraphics[width=.9\columnwidth]{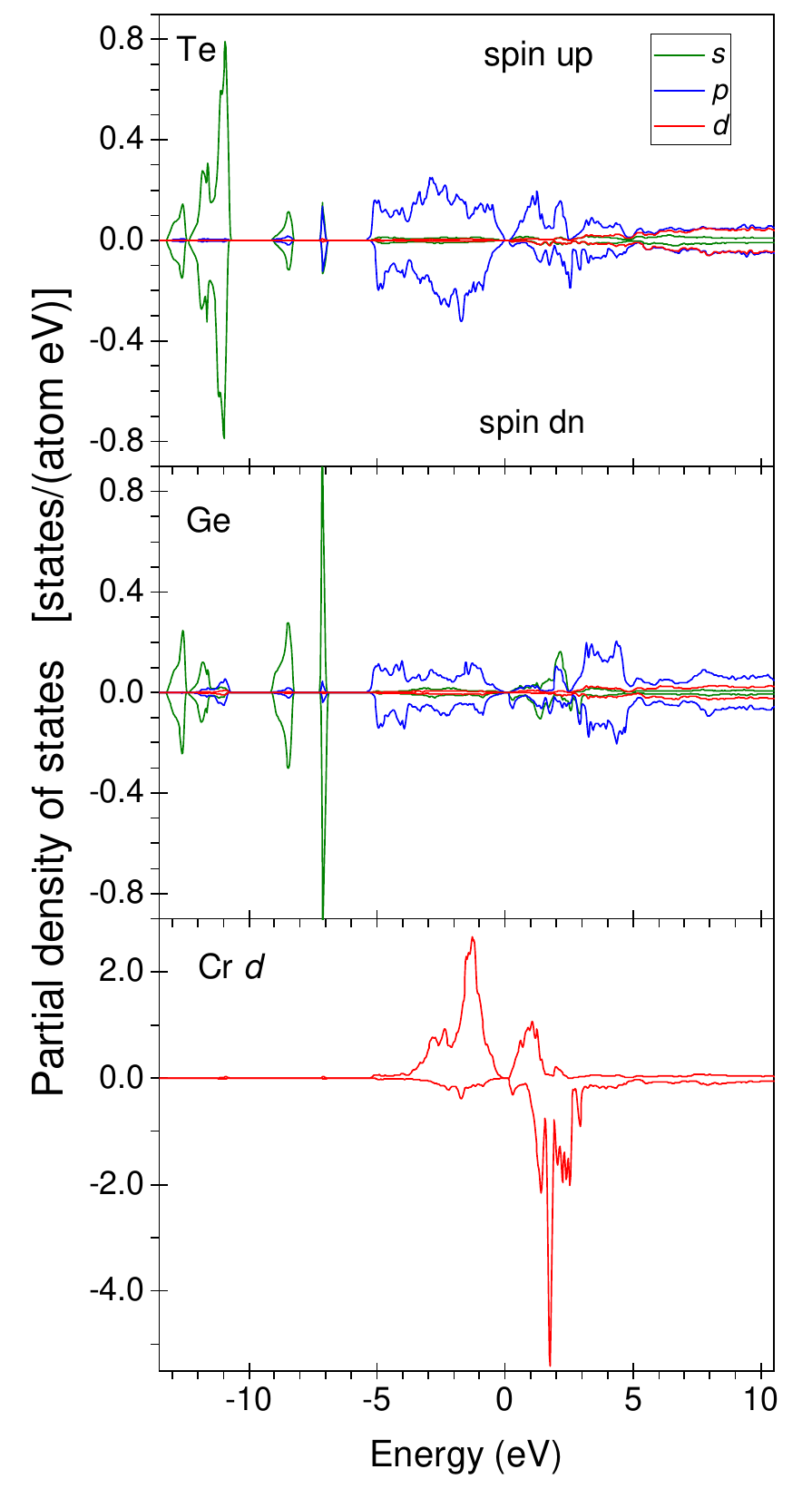}
\end{center}
\caption{\label{PDOS_CGT} The partial DOSs of
 bulk $\cgt$ calculated in the GGA approximation. }
\end{figure}
Figure \ref{PDOS_CGT} presents the partial density of states of $\cgt$
in the GGA approximation. The Te 5$s$ states consist of four peaks and they are located mostly between $-$12.2 and $-$10.7 eV below the Fermi level.
 The Ge 4$s$ states are located mostly
between $-$7.2 and $-$7 eV. Other peaks located at the lower energy
originate from the hybridization with Te 5$s$ states. The 5$p$ states
of Te and Ge are found to be at $-$5.5 eV to 5.0 eV energy interval in
$\cgt$. The spin splitting of the Te and Ge $p$ states is quite
small. The Cr 3$d$ states are situated in the $-$5 eV to 3.2 eV
energy interval. The empty Cr 3$d$ spin-up DOSs present by relatively
weak peak at 0 eV to 1.3 eV interval. The empty Cr 3$d$ spin-down
DOSs consist of three intensive peaks in the 0.5 eV to 3.2 eV interval.

\rtbl{tbl:cgt-bulk-mm} summarizes calculated spin, orbital magnetic
moment and band-gap size.  The trend of magnetic moment change with
$U$-parameter is similar to the trend in the $\cri$ case.  While the
spin moment is decreasing with increasing $U$-parameter in the AMF
scheme, it changes to opposite direction in the FLL scheme.  The
calculated orbital moments in the \rtbl{tbl:cgt-bulk-mm} are too small
to assign physical meaning though the value is getting more negative
with increasing $U$ value.  The measured saturated magnetization is
$\SI{2.92}{\mub}$ at \SI{5}{\K}~\cite{carteaux1995jpcm} and the orbital
moment is $\SI{-0.045}{\mub}$~\cite{kim2019prl}.  The calculated
orbital moment with GGA+$U$ is much smaller than this measured result.

\begin{table}[hbt]
	\caption{On-site spin $\mspin$ and orbital $\morb$ magnetic
          moment of Cr of $\cgt$ bulk. For the FLL scheme calculation,
          the site exchange $J$ is set to be 0 Ry.}
	\label{tbl:cgt-bulk-mm}
	\bgroup
	\def\arraystretch{1.1}
	\begin{tabular*}{\linewidth}{l @{\extracolsep{\fill}} crrrrccc}
		\hline
		\hline
		\\[-1em]
		{$\cgt$} & & $U$ & &  \multicolumn{2}{c}{Cr}  &    &  gap \\
		\\[-1em]  
		\cline{1-1} \cline{3-3} \cline{5-6} \cline{8-8}
		\\[-1.1em]
		Method   & & Ry &  & $\mspin$   & $\morb$    &     & eV \\
		\\[-1.1em]
		\hline                                      
		\\[-1.1em]
		FLAPW & &         &  & 3.06 &  0.004 & & 0.18   \\
		RLMTO & &         &  & 3.34 &  0.029 & & 0.00   \\
                sum rules & &     &  & 3.27 &  0.031 & &    \\
		AMF   & & 0.10    &  & 2.91 &  0.003 & & 0.15   \\
	 	      & & 0.20    &  & 2.65 &  0.002 & & 0.08   \\
		      & & 0.40    &  & 1.93 & -0.003 & & 0.00   \\
		FLL   & & 0.10    &  & 3.21 &  0.001 & & 0.14   \\
		      & & 0.20    &  & 3.36 & -0.001 & & 0.08   \\
		      & & 0.40    &  & 3.60 & -0.003 & & 0.00   \\
		Exp~\cite{kim2019prl} &&    &&  &-0.045&&  \\		    
		\\[-1.0em]
		\hline\hline
	\end{tabular*}
	\egroup
\end{table}	

\subsubsection {The XAS and XMCD spectra}
\begin{figure}[hbt]
	\centering
	\includegraphics[width=1.0\linewidth]{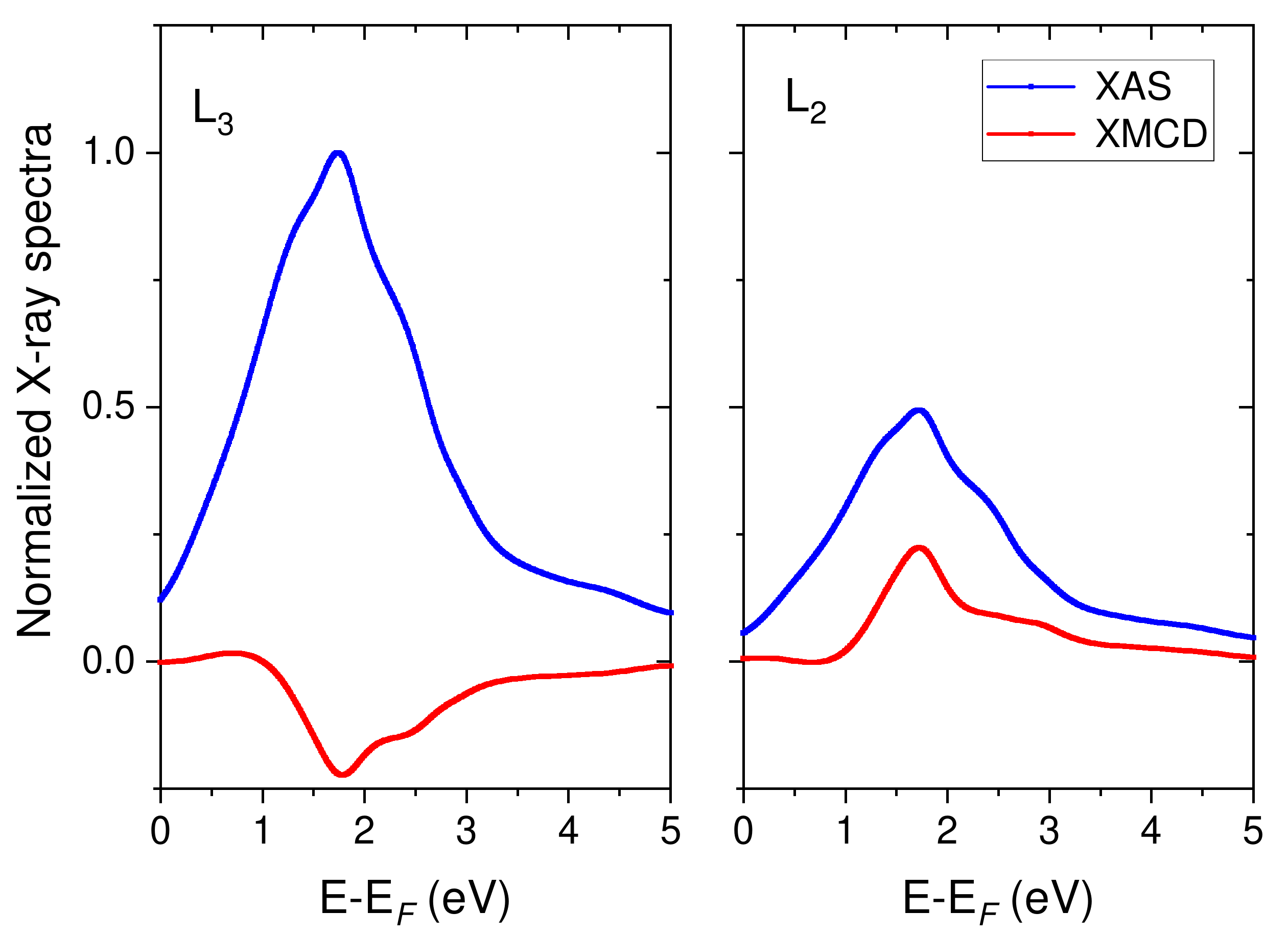}
	\caption{Calculated bulk $\cgt$ XAS (blue line) and XMCD
          (red line) spectra with GGA functional. Left(Right) panel
          shows Cr L$_{3}$ (L$_{2}$) edge spectra. The figures
          includes raw data before adding broadening effects to show
          fine structure change.}
	\label{fig:cgt-bulk-xmcd}	
\end{figure}

\begin{figure}[hbt]
	\centering
	\includegraphics[width=1.1\linewidth]{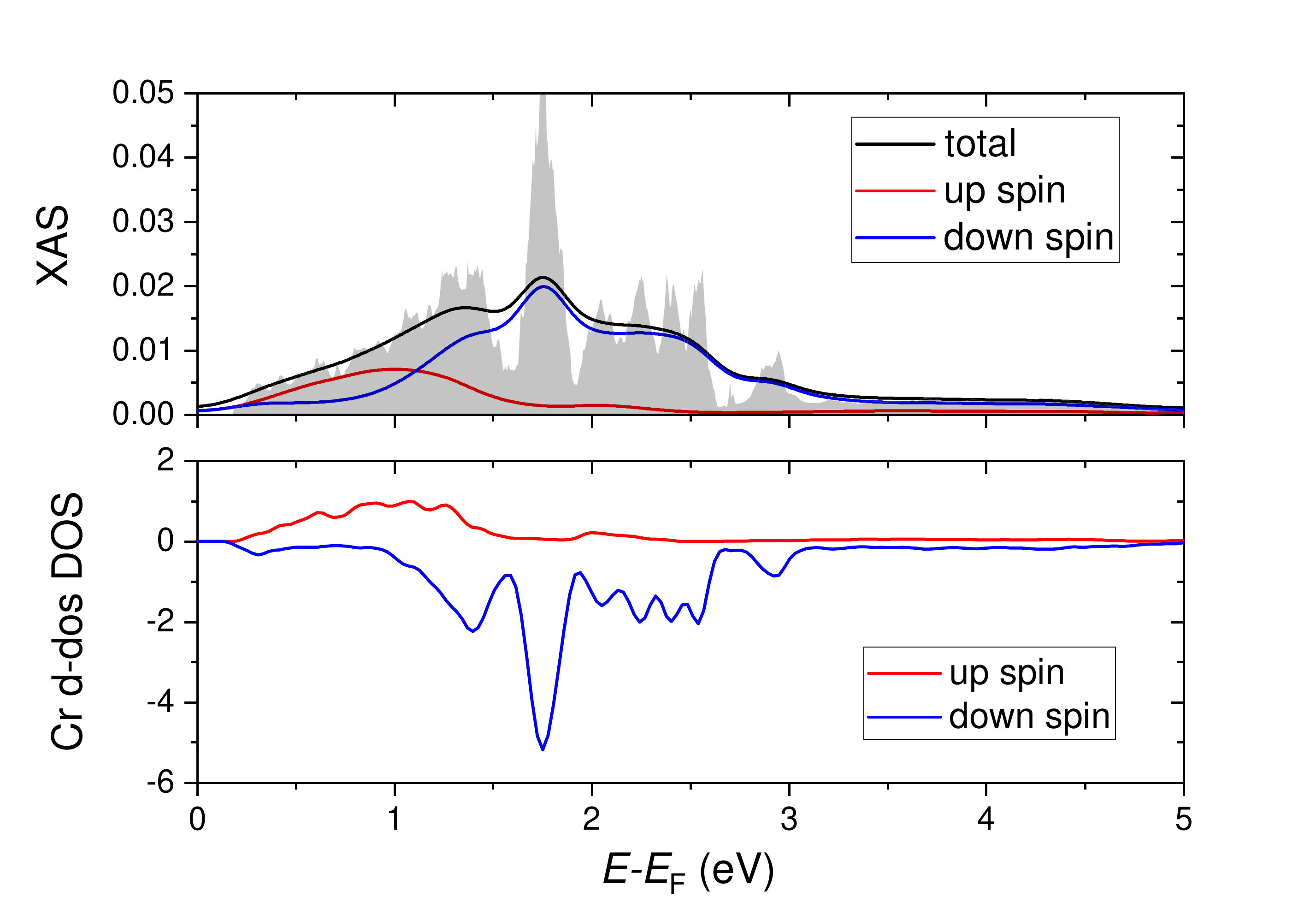}
	\caption{Cr L$_{3}$ edge XAS of $\cgt$ (top panel) and spin
          resolved DOS (bottom panel). Red(Blue) line is up(down) spin
          contribution. The shadow area presents raw spectra.}
	\label{fig:cgt-xas-dos}	
\end{figure}

\rFig{fig:cgt-bulk-xmcd} shows calculated XAS and XMCD spectra with
GGA functionals.  The left panel is Cr-$L_{3}$-edge spectra and the
right panel is Cr L$_{2}$-edge spectra.  Both XASs show three features
which are a strong central peak around 1.8 eV, and two shallow
shoulders on both sides of the central peak around 1.2 eV and 2.3 eV.
Since the distance between peaks is rather close, the spectra look
like a broad peak unlike the Cr L$_3$-edge XAS of $\cri$ which shows
separated peaks in the XMCD spectra.

The top panel of \rfig{fig:cgt-xas-dos} shows Cr L$_{3}$ XAS spectra
which is calculated by using radial matrix elements and the Cr
$d$-state DOS.  It clearly shows that all three features of the XAS
spectra are attributed to the down spin DOS.  The bottom panel shows
the Cr $d$-state DOS which has a broader bandwidth than Cr $d$-state
of $\cri$. It suggests that the Cr atom in $\cgt$ is involved in
stronger hybridization than in $\cri$.  In their atomic calculations
of XAS/XMCD spectra of $\cgt$, Watson $\etal$~\cite{watson2020prb}
observed that a strong hybridization parameter is required to obtain
results that show good agreement with the experimental result.
Menichetti $\etal$~\cite{menichetti20192m} also reported that
increasing the applied $U$ potential decreases the band-gap and attributed
it to strong hybridization between the Cr and Te atoms.

\begin{figure}[htb]
    \centering
    \includegraphics[width=0.95\linewidth]{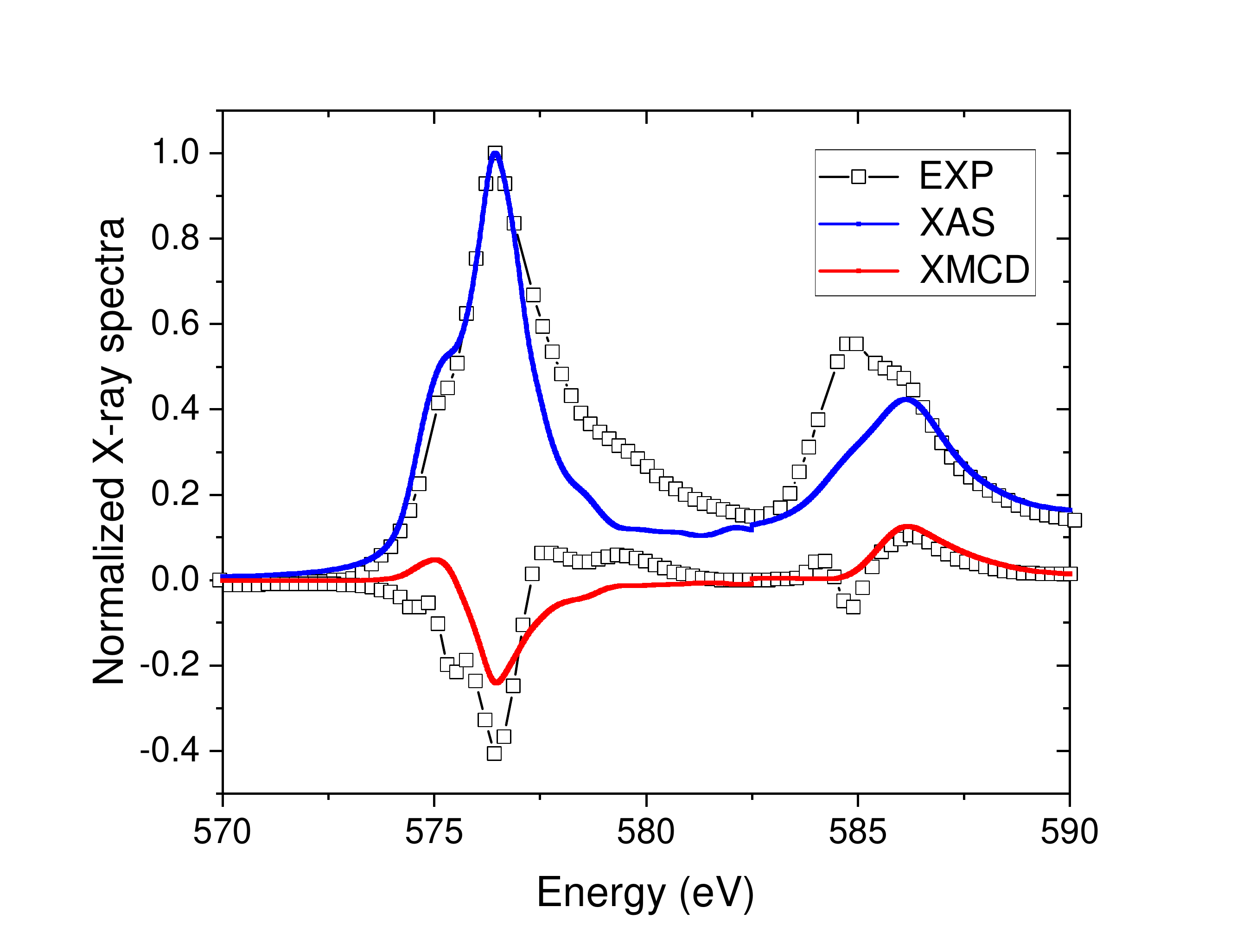}\\	
    \includegraphics[width=0.95\linewidth]{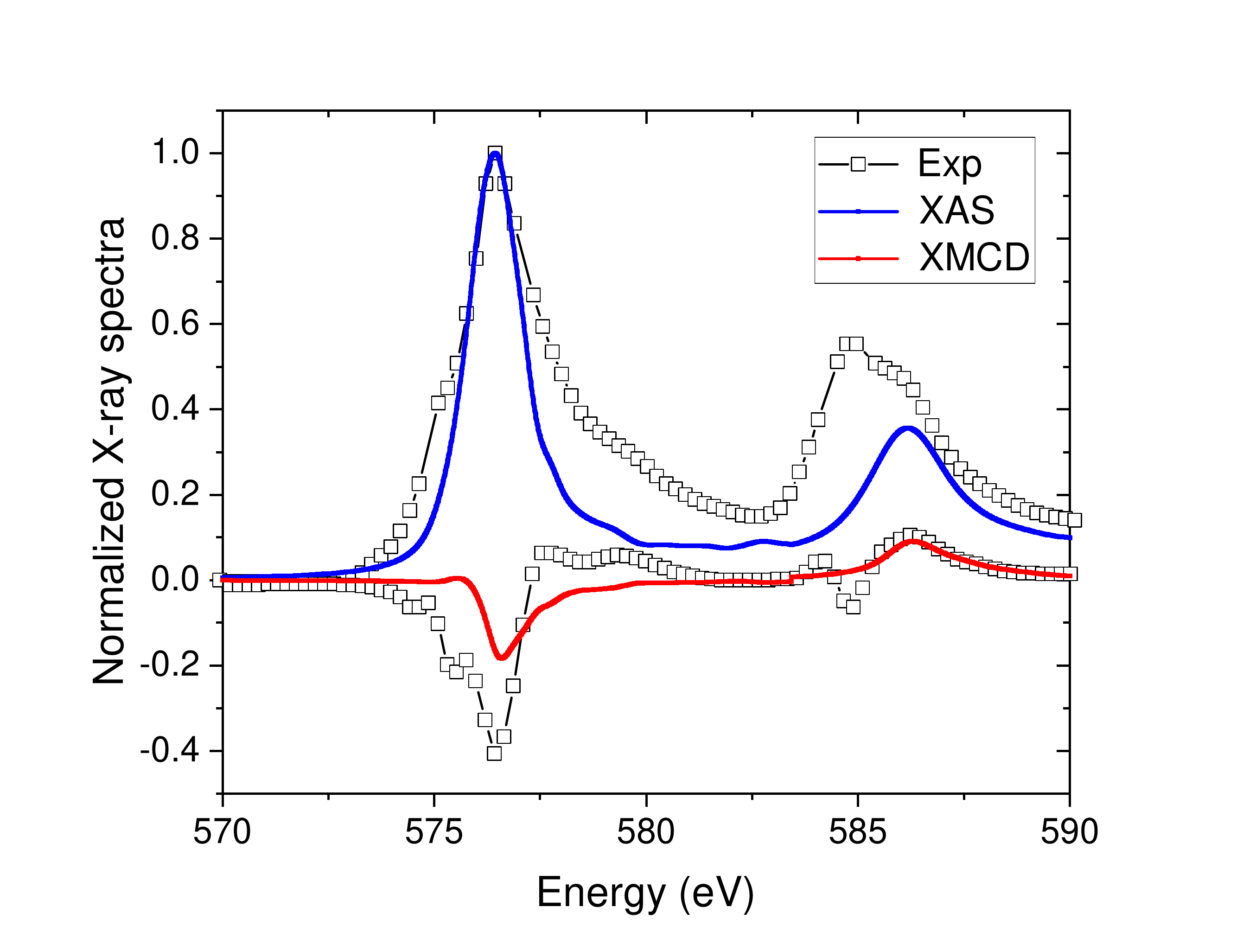}
	\caption{Comparison between calculated and measured Cr L$_{3}$
          and L$_{2}$-edge XAS and XMCD of $\cgt$.  The calculation
          was performed with the GGA+$U$ method. The employed $U$
          value was 0.1 Ry for both the FLL and the AMF scheme. The
          top (bottom) panel show the result with FLL (AMF)
          scheme. The blue (red) line is theoretical XAS
          (XMCD). Experimental data (open squares) were obtained from
          Kim $\etal$~\cite{kim2019prl}.}
	
	\label{fig:cgt-xas-ldau1}
\end{figure}

\rFig{fig:cgt-xas-ldau1} shows the spectra that are calculated with
the FLL(top panel) and the AMF(bottom panel) schemes of the GGA+$U$
method. The top panel shows that two lower energy peaks are separated
further compared to \rfig{fig:cgt-bulk-xmcd}. Although the
L$_{3}$-edge of XAS spectra which is calculated using the FLL scheme
shows good agreement with the experiment, overall agreement between
theoretical and experimental spectra is not so good. Using a higher
$U$ value does not improve the agreement. It seems that the Hubbard
$U$ is less effective for the Cr atom in the $\cgt$ than in the $\cri$
since the Cr atom in the $\cgt$ is involved in stronger hybridization.

\section{conclusion}
We performed first principles electronic structure calculations for
bulk and thin film structures of m2DvdW materials $\fgt$, $\cri$ and
$\cgt$ with the GGA functional and the GGA+$U$ methods. XAS, XMCD
spectra were calculated using wavefunctions that were generated by
first principles calculations. We show that the GGA is applicable
for the metallic $\fgt$ and the GGA+$U$ method with a rather higher $U$
value is required for the semiconducting $\cri$.  Although $\cgt$
is a semiconductor, because of the strong hybridization between Cr and
Te atom, the spectral line shapes are not so sensitive to the values of 
the Hubbard U. The core-hole effects are not so strong to alter the spectral
line shapes. Our calculations have been able to provide help in clarifying the various contributions to specific features of the experimental XMCD measured spectra.  Although the complexity of the X-ray spectra excitations are formidable for calculations, by identifying which features of the spectra are primarily associated with particular elemental electron excitations, more sophisticated electron correlation treatments can be explored and tested with new experiments.

\section*{Acknowledgments}

This work was supported by the U.S.~Department of Energy, Office of
Science, Office of Basic Energy Sciences, Materials Sciences and
Engineering Division, and Early Career Research Program.  Ames
Laboratory is operated for the U.S.~Department of Energy by Iowa State
University under Contract No.~DE-AC02-07CH11358.

\appendix

\bibliography{aaa}

\bigskip



\end{document}